\documentclass[pra,aps,preprint,superscriptaddress,12pt]{revtex4-2}

\usepackage{physics}
\usepackage{xspace}
\usepackage{graphicx}
\usepackage{setspace}

\usepackage[colorlinks, citecolor={blue}, linkcolor={black}, urlcolor={blue}]{hyperref}

\usepackage{titlesec}
\titleformat{\section}{\normalfont\Large\bfseries}{\thesection}{1ex}{}
\titleformat{\subsection}{\normalfont\large\bfseries}{\thesubsection}{1ex}{}
\renewcommand{\thesection}{\arabic{section}}
\renewcommand{\thesubsection}{\arabic{section}.\arabic{subsection}}

\begin{document}

\setstretch{1}
\title{Quantum spin Hall effects in van der Waals materials}

\author{Jian Tang}
\affiliation{Department of Physics, Boston College, Chestnut Hill, MA, USA}

\author{Thomas Siyuan Ding}
\affiliation{Department of Physics, Boston College, Chestnut Hill, MA, USA}

\author{Chengdong Wang}
\affiliation{Department of Materials Science and Engineering, Texas A\&M University, College Station, TX, USA}

\author{Ning Mao}
\affiliation{Max Planck Institute for Chemical Physics of Solids, 01187, Dresden, Germany}

\author{Vsevolod Belosevich}
\affiliation{Department of Physics, Boston College, Chestnut Hill, MA, USA}

\author{Yang Zhang}
\affiliation{Department of Physics and Astronomy, University of Tennessee, Knoxville, TN, USA}
\affiliation{Min H. Kao Department of Electrical Engineering and Computer Science, University of Tennessee, Knoxville, TN, USA}

\author{Xiaofeng Qian}
\affiliation{Department of Materials Science and Engineering, Texas A\&M University, College Station, TX, USA}
\affiliation{Department of Physics and Astronomy, Texas A\&M University, College Station, TX, USA}
\affiliation{Department of Electrical and Computer Engineering, Texas A\&M University, College Station, TX, USA}

\author{Qiong Ma}
\email{maqa@bc.edu}
\affiliation{Department of Physics, Boston College, Chestnut Hill, MA, USA}

\maketitle
\tableofcontents

\newpage
\section*{Abstract}

\textbf{The quantum spin Hall (QSH) effect, first predicted in graphene by Kane and Mele in 2004, has emerged as a prototypical platform for exploring spin-orbit coupling, topology, and electronic interactions. Initially realized experimentally in quantum wells exhibiting characteristic QSH signatures, the field has since expanded with the discovery of van der Waals (vdW) materials. This review focuses on vdW systems, which offer unique advantages: their exposed surfaces enable a combination of surface-sensitive spectroscopic and microscopic tools for comprehensive detection of the QSH state; mechanical stacking with other vdW layers facilitates symmetry engineering and proximity effects; and moiré engineering introduces layer skyrmion topological phases and strong correlation effects. We highlight two monolayer families, 1T$^\prime$-MX$_2$ and MM$^\prime$X$_4$, represented by WTe$_2$ and TaIrTe$_4$, respectively. These materials exhibit QSH phases intertwined with or in close proximity to other quantum phases, such as excitonic insulators, charge density waves, and superconductivity. Their low crystal symmetry and topology enable rich quantum geometrical responses, ranging from nonlinear Hall effects to circular photogalvanic effects. We also discuss moiré systems, which combine topology with flatband physics and enhanced correlations, driving spontaneous symmetry breaking and transitions from QSH to quantum anomalous Hall (QAH) states. Remarkably, fractionalized QAH and QSH states have recently been observed in moiré systems, significantly advancing the field of condensed matter physics. Finally, we explore emerging applications of QSH and derived materials, such as using nonlinear Hall effects for quantum rectification in microwave energy harvesting and harnessing fractional anomalous states for topological quantum computing.} 

\section{Introduction}

\subsection{A brief history of QSH insulator}
The QSH insulator ~\cite{kane2005quantum,bernevig2006quantum,qi2010quantum,maciejko2011quantum}, a two-dimensional (2D) topological insulator (TI), differs fundamentally from conventional insulators. While both exhibiting a bulk electronic gap, a QSH insulator uniquely hosts one-dimensional (1D) metallic edge states. The distinction arises from the bulk band structure: in a conventional insulator, the conduction and valence bands retain their normal ordering, forming a trivial band gap, whereas in a QSH insulator, the conduction and valence bands swap at specific points in the Brillouin zone. This band inversion gives rise to helical edge states with spin-momentum locking, where electrons with opposite spins propagate in opposite directions. These helical edge states are typically protected by time-reversal symmetry, which prevents backscattering from non-magnetic impurities and other time-reversal-symmetric perturbations. This protection ensures robust edge conduction of one conductance quantum per edge, enabling dissipationless transport and making QSH insulators promising candidates for low-energy-consumption electronics.

The QSH insulator was first predicted in graphene, where spin-orbit coupling (SOC) was theorized to gap out the Dirac cones and give rise to helical edges~\cite{kane2005quantum}. However, the intrinsic SOC in graphene is extremely weak due to the lightness of carbon atoms, with a measured strength of only $\sim$42.2 $\mu$eV~\cite{sichau2019resonance}. As a result, experimental observation of the QSH phase in pristine graphene has remained elusive. Nevertheless, helical edge states in graphene have been reported by circumventing its intrinsic band structure and exploiting its Landau levels. For instance, Young et al. utilized the zero Landau level ($\nu = 0$) in monolayer graphene~\cite{young2014tunable, veyrat2020helical}. Specifically, further applying a large in-plane magnetic field drives graphene into a ferromagnetic QSH state, which is equivalent to two copies of quantum Hall edges, protected from mixing by the $U(1)$ symmetry of spin rotations in the plane perpendicular to the magnetic field~\cite{abanin2006spin, kharitonov2016interplay}. Sanchez-Yamagishi et al. further constructed a QSH-like phase in an electron-hole bilayer by stacking two monolayer graphene flakes at a large angle, preventing hybridization between the layers~\cite{sanchez2017helical}. By applying a perpendicular displacement field, the two layers can be tuned to electron and hole fillings, respectively. In this configuration, helical 1D edge conduction is realized, with counterpropagating modes formed by the $\nu = 1$ and $\nu = -1$ Landau levels in the separate electron and hole layers.

The QSH state was later proposed by Bernevig et al.~\cite{bernevig2006quantum} in 2006 in mercury telluride–cadmium telluride (HgTe/CdTe) semiconductor quantum well systems. By tuning the thickness of the HgTe layer, a topological phase transition can be induced via band inversion. This prediction was experimentally confirmed by König et al.~\cite{konig2007quantum} in 2007, who fabricated HgTe/CdTe quantum wells and observed quantized edge conductance at cryogenic temperatures. Subsequently, Knez et al.~\cite{knez2011evidence} reported a QSH state in InAs/GaSb heterostructures, where a type-II band alignment facilitates an inverted band structure. 

\subsection{Recent developments of QSH insulator in vdW materials}

The emergence of a variety of vdW materials beyond graphene has energized the field, offering new avenues for experimental exploration. Theoretically, intrinsic 2D crystals of monolayer 1T$^\prime$-phase transition metal chalcogenides, such as binary compounds MX$_2$ (M = Mo, W; X = S, Se, Te) and ternary compounds MM$^\prime$Te$_4$ (M = Ta, Nb; M$^\prime$ = Ir, Rh), are predicted to be QSH insulators~\cite{Qian2014Quantum,liu2017van}. Additionally, numerous other materials, such as ZrTe$_5$, HfTe$_5$, Bi$_4$Br$_4$, and Bi$_4$I$_4$, have been proposed as QSH insulators~\cite{luo2015room,zhou2014large,Liu2016WeakTI,shumiya2022evidence,autes2016novel,xu2024realization,weng2014transition,wu2016evidence,li2016experimental,tang2021two}. Many of these materials feature large topological gaps, making them promising candidates for high-temperature operation (see Table~\ref{table1} for more details). Interestingly, some of these materials, such as Bi$_4$Br$_4$ and Bi$_4$I$_4$, exhibit quasi-1D structures and can adopt different topological phases, including weak TI or higher-order TI states with helical hinge states, which can be further tuned by strain~\cite{zhou2014large,Liu2016WeakTI,Yoon2020quasione,shumiya2022evidence,autes2016novel}. These materials bridge topology with the rapidly growing 2D vdW material community. The unique advantages of vdW materials, particularly the ability to stack and twist layers to create new structures and engineer novel quantum phases, make them highly versatile for exploring topological phenomena.

Experimentally exploring the transport properties of these monolayers requires significant effort and recipe development, including isolating the monolayer, developing protection methods (as many of these monolayers are air-sensitive), and fabricating ohmic contacts~\cite{fei2017edge, wu2018observation,shi2019imaging,Tang2024DualQSH}. These processes can take years of dedicated work. Fortunately, the exposed surfaces of these materials allow access to powerful spectroscopic and microscopic techniques, such as angle-resolved photoemission spectroscopy (ARPES) and scanning tunneling microscopy/spectroscopy (STM/STS)~\cite{tang2017quantum,autes2016novel}. ARPES provides direct insight into the spectral function and topological gap, and its nano- or micro-scale versions offer spatial resolution to possibly distinguish 2D bulk from edge states. STM/STS enables energy-resolved measurements of the local density of states with atomic resolution, providing simultaneous access to both bulk and edge electronic structures in real space, and is compatible with magnetic field. These techniques can be readily applied to samples prepared via a range of methods, including mechanical exfoliation and epitaxial growth on substrates. For instance, recent STM work by Shumiya et al.~\cite{shumiya2022evidence} identified $\alpha$-Bi$_4$Br$_4$ as a room-temperature QSH insulator with a large gap of $\sim$260 meV. Similarly, Xu et al.~\cite{xu2024realization} examined monolayer ZrTe$_5$ grown on bilayer graphene/SiC substrates, identifying a gap of 254 meV and edge density of states.

Monolayer WTe$_2$ and TaIrTe$_4$ have been fabricated into high-quality dual-gated electronic devices, enabling systematic transport studies~\cite{fei2017edge, wu2018observation, Tang2024DualQSH}. These studies have demonstrated enhanced nonlocal responses and quantized edge conductance inside the QSH gap. More intriguingly, these materials exhibit not only the QSH effect but also new phases intertwined with topology. In monolayer WTe$_2$, the topological gap is believed to be facilitated by exciton condensation~\cite{jia2022evidence,sun2022evidence,pereira2022topological,wu2024charge}, while in TaIrTe$_4$, enhanced density of states near van Hove singularities leads to a correlated topological gap at finite doping, making it a dual QSH insulator~\cite{Tang2024DualQSH,liu2024prediction}. Additionally, their low crystal symmetry and inverted band structures result in large Berry curvature and rich quantum geometrical properties.

Moiré engineering introduces a new family of 2D topological insulators, leveraging layer pseudospin structures~\cite{wu2018hubbard,wu2019topological,devakul2021magic,reddy2023fractional,xu2024maximally,wang2024fractional}. Specifically, moiré systems combine topology with flatband physics and enhanced correlations, leading to spontaneous symmetry breaking and transitions from QSH to QAH states. Remarkably, fractionalized versions of QAH and QSH have recently been reported, significantly advancing the field~\cite{cai2023signatures,zeng2023thermodynamic,park2023observation,xu2023observation,kang2024evidence}.

Furthermore, the stacking flexibility of vdW materials allows integration with magnetic or superconducting materials to engineer symmetry and proximity effects. Integration with 2D magnetic and superconducting materials enables novel quantum phenomena crucial for spintronics and quantum computing. For instance, magnetic proximity effects in WTe$_2$-based heterostructures with CrI$_3$ and Cr$_2$Ge$_2$Te$_6$ modulate edge conductance and induce spin-polarized transport~\cite{zhao2020magnetic}, while WTe$_2$/Fe$_3$GeTe$_2$ interfaces stabilize Néel-type skyrmions via Dzyaloshinskii-Moriya interactions~\cite{wu2020neel}. Superconducting proximity in WTe$_2$/NbSe$_2$ heterostructures enables potential topological superconductivity, perhaps supporting fault-tolerant quantum computing~\cite{lupke2020proximity}. Additionally, unconventional spin-orbit torques in WTe$_2$ enable field-free deterministic magnetization switching in magnetic materials~\cite{macneill2017control,kao2022deterministic, liu2023field,zhang2023room,bainsla2024large}, advancing energy-efficient spintronic memory.

The QSH insulator could enable a wide range of applications in energy-efficient electronics and spintronics. Field-tunable topological phase transitions in QSH materials allow for dissipationless transistors~\cite{Qian2014Quantum}, while strain engineering provides an alternative control for topological phase transition~\cite{liu2014manipulating}. QSH edge states facilitate efficient spin-charge conversion~\cite{kondou2016fermi}, which is crucial for spintronics. We will also discuss that the nonlinear Hall (NLH) effect in QSH insulator-derived materials presents a novel mechanism for energy harvesting, which may outperform conventional rectifiers due to its unique Hall-based rectification~\cite{onishi2024high}. Further research is needed to optimize NLH-based energy harvesting for practical applications. Moreover, we discuss how to utilize the recently discovered fractional states for fault-tolerant quantum computing. 

\begin{table}
\centering
\begin{tabular}{ |p{4.5cm}|p{2.6cm}|p{4.1cm}|p{4.1cm}| }
\hline
Materials & \begin{tabular}{@{}c@{}}Gap formation \\mechanism\end{tabular} & \begin{tabular}{@{}c@{}} Theory \\ Gap size (meV) \end{tabular} & \begin{tabular}{@{}c@{}} Experiment \\ Gap size (meV) \end{tabular} \\
\hline
Graphene~\cite{gmitra2009band,boettger2007first} & \begin{tabular}{@{}c@{}}Type 1\end{tabular}  & $0.001 \sim 0.050$ & / \\
\hline
\begin{tabular}{@{}c@{}} 1T$^{\prime}$-MX$_2$ \\ (M= Mo, W; \\ X=S, Se, Te) \\ \cite{Qian2014Quantum,ugeda2018observation,chen2018large,zheng2016quantum,fei2017edge,tang2017quantum, Chen2020WSe2, wang2021landau}  \end{tabular}  & \begin{tabular}{@{}c@{}}Type 1\end{tabular}  & \begin{tabular}{@{}c@{}}$45\sim76$ (MoS$_2$) \\ $31\sim88$ (MoSe$_2$) \\$44\sim110$ (WS$_2$) \\$36\sim116$ (WSe$_2$) \\$-262\sim-300$ (MoTe$_2$) \\ $-122 \sim -131$ (WTe$_2$) \\ $100\sim141$ (WTe$_2$) \end{tabular} & \begin{tabular}{@{}c@{}} $\sim 120$, $129$ (WSe$_2$) \\ $5$, $56\pm14$, $60$ (WTe$_2$) \\ $ $  \end{tabular}  \\
\hline
\begin{tabular}{@{}c@{}}MM$^{\prime}$Te$_4$ \\ (M= Ta, Nb; \\ M$^{\prime}$=Ir, Rh)  \cite{liu2017van, Tang2024DualQSH} \end{tabular} & \begin{tabular}{@{}c@{}}Type 1\end{tabular}   & \begin{tabular}{@{}c@{}}$\sim23$ (TaIrTe$_4$) \\ $\sim30$ (NbIrTe$_4$)  \\$\sim58$ (TaRhTe$_4$)  \\$\sim65$ (NbRhTe$_4$) \end{tabular} & $\sim27$ (TaIrTe$_4$) \\
\hline
\begin{tabular}{@{}c@{}}SnX \\ (X = F, Cl, Br, \\ I, OH, CN) \\
\cite{Xu2013Z2SnX,Ji2015SnCN} \end{tabular} & Type 2 & 250-350 & /  \\
\hline
\begin{tabular}{@{}c@{}} MTe$_5$  \\ (M=Zr, Hf) \cite{weng2014transition,wu2016evidence,li2016experimental,tang2021two,xu2024realization} \end{tabular} & \begin{tabular}{@{}c@{}}Type 1\end{tabular}   & \begin{tabular}{@{}c@{}}$\sim100$ (HfTe$_5$)   \\$\sim400$ (ZrTe$_5$) \end{tabular}  & $80\sim254$, 0.28 (ZrTe$_5$)\\
\hline
Sb element layer \cite{zhang2012topological} & \begin{tabular}{@{}c@{}}Type 1\end{tabular} & 500 & / \\
\hline
Bi element layer \cite{murakami2006quantum,reis2017bismuthene,Zhao2025realization} & \begin{tabular}{@{}c@{}}Type 1\end{tabular}   & 250 & 800 \\
\hline
\begin{tabular}{@{}c@{}} MBi \\ (M=B, Al, Ga, \\ In, Tl) \cite{chuang2014prediction} \end{tabular}  &  \begin{tabular}{@{}c@{}}Type 2\end{tabular}   & 250-560 & / \\
\hline
\begin{tabular}{@{}c@{}} MX \\ (M = Zr, Hf; \\ X = Cl, Br, I) \cite{zhou2015new} \end{tabular}  &  \begin{tabular}{@{}c@{}}Type 1\end{tabular}   & 100-400 & / \\
\hline
\begin{tabular}{@{}c@{}} BiX/SbX \\ (X= H, F, Cl, Br) \cite{song2014quantum} \end{tabular}  &  \begin{tabular}{@{}c@{}}Type 1\end{tabular}   & 300-1000 & / \\
\hline
\begin{tabular}{@{}c@{}} Bi$_4$X$_4$ \\ (X=F, Br, I) \cite{luo2015room,zhou2014large,Liu2016WeakTI,shumiya2022evidence,autes2016novel}  \end{tabular}  &  \begin{tabular}{@{}c@{}}Type 1\end{tabular}   & \begin{tabular}{@{}c@{}} $ 700$ (Bi$_4$F$_4$) \\$> 200$ (Bi$_4$Br$_4$)  \\ $37\sim158$ (Bi$_4$I$_4$) \end{tabular}& \begin{tabular}{@{}c@{}} $\sim 50$ (Bi$_4$I$_4$) \end{tabular}   \\
\hline
\begin{tabular}{@{}c@{}}Complex ternary \\TaCX \\ (X = Cl, Br, I) \cite{zhou2016two} \end{tabular} & \begin{tabular}{@{}c@{}}Type 1\end{tabular}   & 300 & /  \\
\hline  
\begin{tabular}{@{}c@{}} Dumbbell Stanene \cite{Tang2014DB_stanene} \end{tabular} & \begin{tabular}{@{}c@{}}Type 2\end{tabular}   & 40 & /  \\
\hline
\end{tabular}
\caption{\textbf{Recent vdW QSH materials.} This table primarily summarizes vdW materials but also includes some materials not strictly defined as vdW layers, serving as examples of distinct band inversion scenarios, as discussed below. Type 1 refers to the band inversion arising from the crystal structure and orbital interactions, while Type 2 refers to the band inversion due to SOC.}
\label{table1}
\end{table}

\section{Design and discovery for QSH insulators}

\subsection{Theoretical discovery of QSH insulators}

\begin{figure}
\centering
\includegraphics[width=\textwidth]{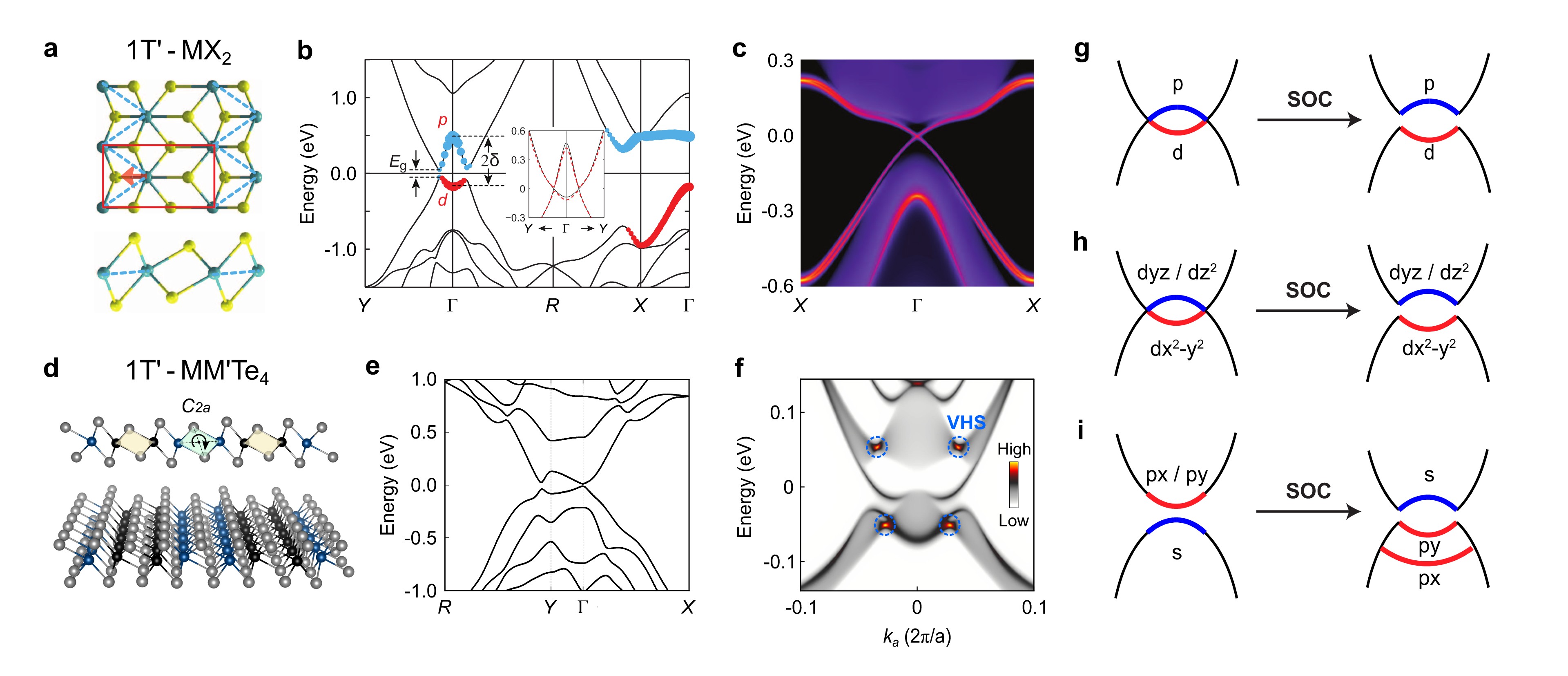}
\caption{\textbf{First-principles prediction and design of QSH insulators.} \textbf{a,} Crystal structure of QSH insulators in 1T$^{\prime}$-MX$_2$ (M=W, Mo; X=Te, Se, S) \textbf{b,} Electronic band structure of 1T$^{\prime}$-MX$_2$ with 1T$^{\prime}$-MoS$_2$ as an example, where the Peierls distortion lowers M's $d$ orbital than Te's $p$ orbital and induces band inversion and the SOC opens the gap at the band crossings. \textbf{c,} Time-reversal pairs of topologically protected edge states in 1T$^{\prime}$-MoS$_2$. \textbf{d,} Crystal structure of QSH insulators in 1T$^{\prime}$-MM$^{\prime}$Te$_4$ (M=Ta, Nb; M$^{\prime}$=Ir, Rh). \textbf{e,} Electronic band structure of 1T$^{\prime}$-MM$^{\prime}$Te$_4$ with 1T$^{\prime}$-TaIrTe$_4$ as an example, where the Peierls distortion lowers Ta's $d_{x^2-y^2}$ orbital than Ta's $d_{yz}$ and $d_{z^2}$ orbitals and induces band inversion and the SOC opens the gap at the band crossings. \textbf{f,} Time-reversal pairs of topologically protected edge states in 1T$^{\prime}$-TaIrTe$_4$. \textbf{g,h} Schematic of the band inversion in 1T$^{\prime}$-MX$_2$ and 1T$^{\prime}$-MM$^{\prime}$Te$_4$, respectively, driven by the interplay of crystal structure and orbital interaction in the absence of SOC, while SOC opens the gap at the band crossings. \textbf{i,} Schematic of the band inversion driven by the interplay of crystal structure, orbital interaction, and SOC, while the system would not exhibit band inversion in the absence of SOC.
Note that Figs.~\textbf{(a-c)} are adapted from Ref.~\cite{Qian2014Quantum}, Figs.~\textbf{(d-f)} from Ref.~\cite{Tang2024DualQSH}.
}
\label{DFT_QSH}
\end{figure}

The design and discovery of QSH insulators~\cite{kane2005quantum,bernevig2006quantum,qi2010quantum,maciejko2011quantum} in 2D materials are guided by a combination of structural, chemical, orbital, and SOC. Generally, QSH insulators can be classified into two categories according to the origin of band inversion -- (a) QSH insulators with band inversion driven purely by the underlying crystal structure and orbital interaction, even in the absence of SOC (Figs.~\ref{DFT_QSH}\textbf{g,h}), (b) QSH insulators with band inversion driven by SOC, that is, behaving as trivial insulators without SOC but becoming topologically nontrivial once SOC is turned on (Fig.~\ref{theory_twist}\textbf{i}).

There are several examples in the first category, driven by crystal structure and orbital interactions, including graphene~\cite{kane2005quantum}, 2D transition metal chalcogenides and their derivatives such as 1T$^{\prime}$-MX$_2$ compounds~\cite{Qian2014Quantum}, MM$^{\prime}$Te$_4$ materials~\cite{liu2017van}, Jacutingaite (Pt$_2$HgSe$_3$)~\cite{Marrazzo2018QSH} and Pd$_2$HgSe$_3$~\cite{Marrazzo2019Z2}, among others. In the 1T$^{\prime}$-MX$_2$ family (where M = Mo, W and X = S, Se, Te), a Peierls distortion and dimerization of the metal chain along one of the in-plane directions alter the crystal potential (Fig.~\ref{DFT_QSH}\textbf{a}), thereby lowering the energy of transition metal $d$-orbitals and driving a band inversion between the transition metal $d$-orbitals and the chalcogen $p$-orbitals near the $\Gamma$ point (Fig.~\ref{DFT_QSH}\textbf{b})~\cite{Qian2014Quantum}. This band inversion — a hallmark feature required for realizing QSH insulator phases — gives rise to Dirac-like band crossings along the high-symmetry Y-$\Gamma$-Y line even before SOC is considered. Once SOC is introduced, a topological gap is opened at these band crossings, stabilizing a $\mathcal{Z}_2 = 1$ QSH insulator phase with robust helical edge states (Fig.~\ref{DFT_QSH}\textbf{c}). It demonstrates the combination of structural engineering (e.g. via 1T$^{\prime}$ distortion) and orbital design (e.g. $p$-$d$ inversion) as an effective strategy for achieving QSH insulators. A similar but more flexible design framework emerges in the ternary transition metal chalcogenide 1T$^{\prime}$-MM$^{\prime}$Te$_4$ (M = Nb, Ta and M$^{\prime}$ = Ir, Rh), where the additional compositional degree of freedom further expands the topological landscape. In these QSH insulators, the design principle is based on engineering interlayer interactions and inversion asymmetry in addition to intralayer orbital hybridization. In the monolayer form, 1T$^{\prime}$-MM$^{\prime}$Te$_4$ exhibits a distorted structure similar to 1T$^{\prime}$-MX$_2$ (Fig.~\ref{DFT_QSH}\textbf{d}), again leading to a $d$-$d$ band inversion driven by the relative shifts of transition metal $d_{x^2-y^2}$, $d_{yz}$, and $d_{z^2}$ orbitals (Fig.~\ref{DFT_QSH}\textbf{e})~\cite{liu2017van}. These structural distortions and orbital interactions drive the formation of Dirac-like points near the Fermi level, which are gapped by SOC into a QSH insulator phase (Fig.~\ref{DFT_QSH}\textbf{f}).

For the second type of QSH insulators, their topological phase only emerges when SOC is turned on~\cite{Xu2013Z2SnX,Ji2015SnCN}. For example, without SOC, stanene (2D Sn) and decorated stanene (SnX, with X=F, Cl, Br, I, OH, CN) behave as a zero-gap semimetal, where the valence and conduction bands -- primarily composed of tin $p_{x,y}$ and $s$ orbitals, respectively -- touch at the $\Gamma$ point. When SOC is introduced, the degeneracy between the $p_x$ and $p_y$ orbitals is lifted, which opens a finite gap at the band crossing where the $p$- and $s$-derived bands are inverted. This SOC-induced band inversion drives the system into a nontrivial QSH insulator with $\mathcal{Z}_2 = 1$, as confirmed by the parity eigenvalues at time-reversal-invariant momenta (TRIM) points.

Recent studies have identified excitonic topological insulator state in monolayer WTe$_2$~\cite{Varsano2020excitonicTI,jia2022evidence,sun2022evidence}, a charge density wave–induced topological insulator state in monolayer TaIrTe$_4$~\cite{Tang2024DualQSH}, and layer skyrmion topological states in moiré transition metal dichalcogenides (TMDs)~\cite{wu2019topological}. These phases will be discussed in detail in later sections.

The above rich mechanisms provide crucial insights for discovering and designing novel QSH insulators across a wide range of 2D materials. By engineering structural distortion, tuning chemical composition or layer degree of freedom, and leveraging SOC, one may control the relative orbital energy levels to induce the band inversion and tailor the gap size of novel QSH insulators, particularly important for practical applications in topological electronics and topological quantum computing.

\subsection{Computational methods}

QSH insulators are characterized by nontrivial \textit{topological invariant} with $\mathcal{Z}_2 = 1$. The principle of bulk-boundary correspondence ensures that a nontrivial bulk topology, indicated by a nonzero topological invariant, guarantees the presence of protected edge states at the system’s boundary. As a result, robust gapless edge states are present in the edge of QSH insulators protected by time-reversal symmetry~\cite{Kane2005z}. These edge states, known as Kramers pairs, exhibit spin-momentum locking, where electrons with opposite spins propagate in opposite directions along the edges. These edge states are robust against perturbations as long as the time-reversal symmetry is preserved, the band remains inverted, and the gap remains open, making QSH insulators an important class of materials for potential applications in topological electronics and topological quantum computing. It is therefore highly desirable to computationally calculate the topological $\mathcal{Z}_2$ invariant for potential QSH insulator candidates and provide the screened candidates for experimental validations. Several first-principles calculation methods have been developed to compute the topological invariants, including TRIM method~\cite{Fu2007TRIM}, Wannier Charge Center (WCC) method ~\cite{Yu2011Wilsonloop,Soluyanov2011WCCs,Taherinejad2014WCCs}, and $n$-field method~\cite{Fukui2005nfield,Fukui2007nfield}.

The TRIM method proposed by Fu and Kane~\cite{Fu2007TRIM} is an efficient approach for determining the $\mathcal{Z}_2$ topological invariant in centrosymmetric materials by using the underlying inversion symmetry and analyzing parity eigenvalues at TRIMs in the Brillouin zone. There are four TRIM points in 2D systems and eight TRIM points in 3D systems. The method involves identifying the TRIM points, computing the parity eigenvalues of the occupied bands, and applying the Fu-Kane formula, given by $(-1)^{\mathcal{Z}_2} = \prod_{i=1}^{N} \prod_{m} \xi_{m}(\Gamma_i)$,
where the product runs over all occupied bands. If $\mathcal{Z}_2 = 1$, the system is in a nontrivial topological phase, while if $\mathcal{Z}_2 = 0$, the system is a trivial or normal insulator. This approach has been widely applied to topological insulators, such as 1T$^{\prime}$-MX$_2$~\cite{Qian2014Quantum} compounds and HgTe/CdTe quantum wells, which exhibit strong SOC and band inversion. Although the TRIM method is limited to centrosymmetric systems, it remains a powerful tool for identifying QSH insulators and guiding the discovery of new topological materials.

The $n$-field method is also based on the formulation of $\mathcal{Z}$ invariant by Fu and Kane~\cite{Fu2006}, \emph{i.e.}, $\mathcal{Z}_2 = \frac{1}{2\pi i} \left[ \oint_{\partial \mathcal{B}^-} A - \int_{\mathcal{B}^-} F \right]$ with $\mathcal{B}^- = [-\pi, \pi] \otimes [-\pi, 0]$. Here, $A$ is Berry gauge potential, and $F$ is the associated Berry field strength, with $A = \text{Tr} \, \psi^{\dagger} \, \mathrm{d} \psi$ and $F = \mathrm{d}A$. The $n$-field method involves discretizing the first Brillouin zone into small plaquettes, computing the associated field strength in each plaquette, and summing over the vortices from the plaquettes for the half of the Brillouin zone, which provides $\mathcal{Z}$ invariant. The $n$-field method can be applied to both centrosymmetric and noncentrosymmetric materials. For example, it has been used to confirm the nontrivial QSH insulator in 1T$^{\prime}$-MX$_2$ (M = Mo, W and X = S, Se, Te)~\cite{Qian2014Quantum} and ternary transition metal chalcogenide 1T$^{\prime}$-MM$^{\prime}$Te$_4$ (M = Nb, Ta and M$^{\prime}$ = Ir, Rh)~\cite{liu2017van}.

The WCC~\cite{Yu2011Wilsonloop,Soluyanov2011WCCs,Taherinejad2014WCCs} method is another powerful approach for calculating topological invariants for both centrosymmetric and noncentrosymmetric materials. In a periodic crystal, the electronic states are described by Bloch wave functions $\psi_{n\mathbf{k}}(\mathbf{r}) = e^{i\mathbf{k} \cdot \mathbf{r}} u_{n\mathbf{k}}(\mathbf{r})$, where $n$ is the band index, $\mathbf{k}$ is the crystal momentum, and $u_{n\mathbf{k}}(\mathbf{r})$ is the cell-periodic part. Wannier functions are then constructed by performing a Fourier transformation of the Bloch states $|W_{n\mathbf{R}}\rangle = \frac{1}{(2\pi)^3} \int_{\text{BZ}} d\mathbf{k} \, e^{-i\mathbf{k} \cdot \mathbf{R}} |u_{n\mathbf{k}}\rangle$, where $|W_{n\mathbf{R}}\rangle$ represents a Wannier function localized at lattice site $\mathbf{R}$. The above single band Wannier function can be extended into multi-band Wannier functions by using a unitary matrix $U_{mn}(\mathbf{k})$ to achieve better localization and chemical intuition, for examples, maximally localized Wannier functions~\cite{Marzari1997MLWF,Souza2001MLWF,Marzari2012MLWF} and quasiatomic orbitals~\cite{Lu2004QUAMBO,Qian2008QO}. To analyze topological properties, hybrid Wannier functions (HWFs) are introduced via partial Wannier transformation, given by $|W_{n l_z} (k_x, k_y)\rangle = \frac{1}{2\pi} \int dk_z \, e^{ik \cdot (r - l_z \hat{c}_z)} |u_{n} (k_x, k_y, k_z)\rangle$, where $l_z$ is the layer index along the localized direction, ($k_x$, $k_y$) remain good quantum numbers, and $\hat{c}_z$ represents the unit vector along the $z$-axis. The center of charge of these HWFs along the Wannierized direction defines WCCs, which are used to determine the topological invariants of the system. By tracking the evolution of WCCs across the Brillouin zone, one can extract $\mathcal{Z}_2$ topological invariants and classify materials as trivial or topological insulators. To determine the $\mathcal{Z}_2$ topological invariants, the connectivity of the WCCs is analyzed. If the WCCs exhibit an even number of crossings along a path connecting two TRIM, the system is classified as a trivial insulator. Conversely, if there is an odd number of crossings, the system is a QSH insulator. 

\subsection{Stacking and moir\'e engineering of QSH insulator}

Over the past few years, stacking and moiré engineering have emerged as powerful tools for designing novel condensed matter phenomena, ranging from superconductivity and magnetism to correlated insulators and ferroelectricity. This approach has also proven highly effective in constructing new QSH insulators, particularly in semiconducting TMD moiré heterostructures. In this section, we explore the theoretical considerations of how topology arises in these heterostructures, even when the individual TMD layers are topologically trivial. Notably, the flat bands of moiré superlattices introduce strong correlation effects, leading to emergent magnetism, the QAH effect, and even fractional QSH and fractional QAH insulators. The experimental progress in these areas will be summarized in the following sections.

\begin{figure}
\centering
\includegraphics[width=0.7\textwidth]{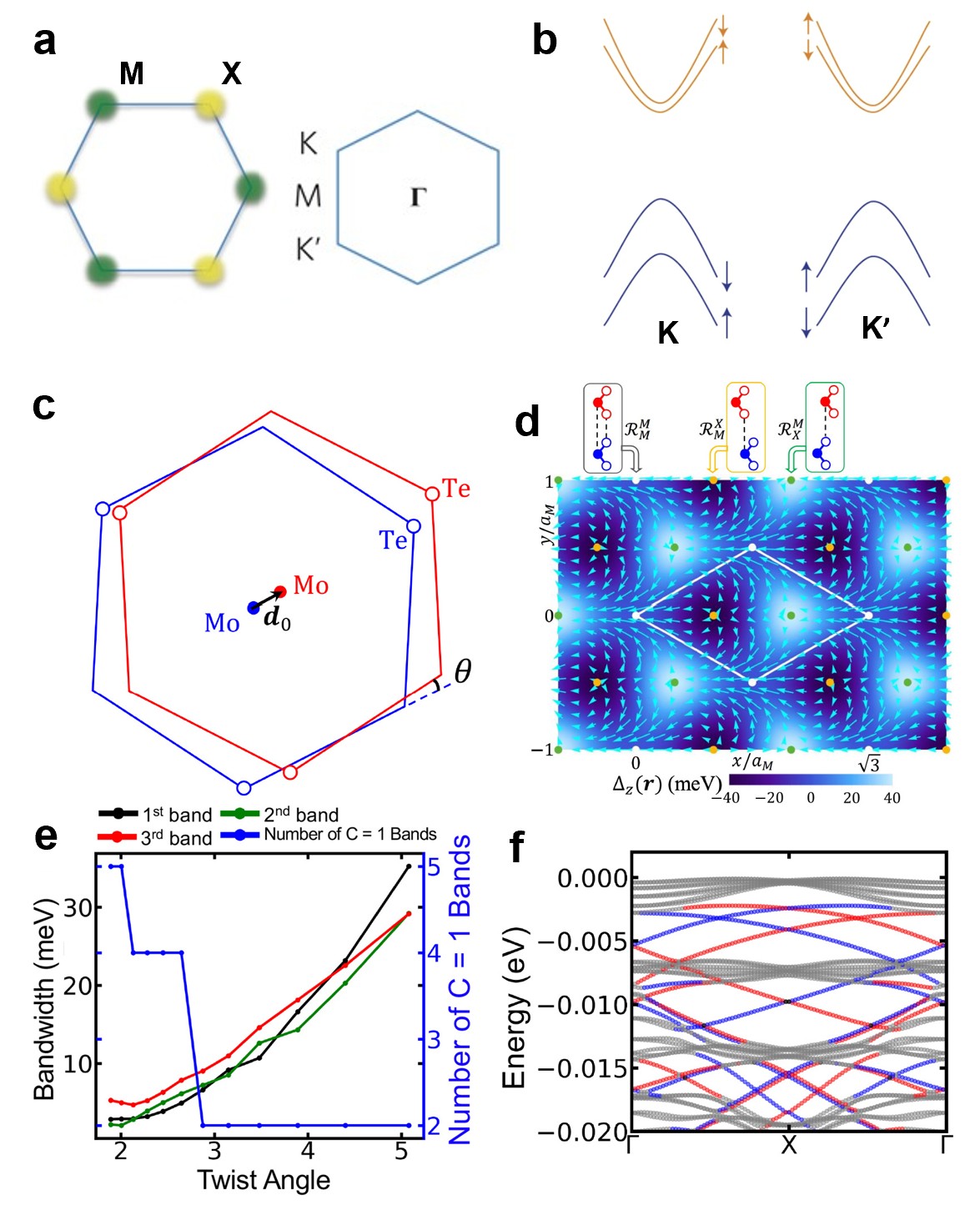}
\caption{\textbf{Theoretical predictions for QSH phase in twisted TMD systems.}  
\textbf{a,} Honeycomb lattice structure of monolayer MX$_2$ (M=W, Mo; X=Te, Se, S) and the first Brillouin zone with high-symmetry points.  
\textbf{b,} Electronic band structure around the $K$ and $K'$ points. The spin (up and down arrows) and valley ($K$ and $K'$) degrees of freedom are locked together.
\textbf{c,} Schematic illustration of twisted MoTe$_2$ with a twist angle $\theta$. $d_0$ denotes the in-plane displacement between the top and bottom Mo atoms. 
\textbf{d,} Real-space moir\'e pattern formed by two twisted TMD layers. The color map shows the variation of $\Delta z$, and the arrows indicate $\Delta x,y$. 
\textbf{e,} Angle-dependent bandwidth and number of $C=1$ bands from $1.89^\circ$ to $5.08^\circ$, calculated by local-basis DFT. The black, green, and red dotted lines represent the bandwidth of the first, second, and third bands, respectively. The blue dotted line indicates the number of $C=1$ bands. 
\textbf{f,} Electronic band structure of a twisted MoTe$_2$ nanoribbon calculated through the DFT Hamiltonian. Trivial bulk states are shown in gray, while spin-up and spin-down edge states are shown in red and blue, respectively.
Note that Figs.~\textbf{(a-b)} are adapted from Ref.~\cite{mak2012control}, Figs.~\textbf{(c-d)} from Ref.~\cite{wu2019topological}, Figs.~\textbf{(e-f)} from Ref.~\cite{xu2025multiple}.
}
\label{theory_twist}
\end{figure}

Monolayer TMDs, such as MoTe$_2$ and WSe$_2$, exhibit unique electronic characteristics originating from the strong SOC. The lack of inversion symmetry in these materials leads to a direct band gap at the $K$ and $K'$ points, accompanied by significant spin splitting in the valence band (Figs.~\ref{theory_twist}\textbf{a-b})~\cite{mak2012control}. Furthermore, due to time-reversal symmetry, if a band near the $K$ point is spin-up, then its counterpart near the $K'$ point must be spin-down, and vice versa. This phenomenon, known as spin-valley locking, effectively renders the SOC near these valleys Ising-type, thereby ensuring spin conservation in the vicinity of the valleys. When two same or different TMD monolayers are stacked together to form a moir\'e superlattice, the resulting stacking pattern induces local rearrangements of atomic positions via lattice relaxation (Figs.~\ref{theory_twist}\textbf{c-d}). As a result, the interlayer distance increases in the MM region while it decreases in the MX/XM regions. This relaxation effect makes the moir\'e bands flatter, which in turn enhances electron-electron interactions~\cite{jia2024moire,zhang2024polarization,mao2024transfer}. 

The earliest theoretical investigations of twisted TMD systems were conducted by Wu et al.~\cite{wu2018hubbard,wu2019topological}, who employed a continuum model to predict various correlated and topological phases, including potential single and double QSH states. Subsequent first-principles calculations performed by multiple research groups~\cite{devakul2021magic,xu2024maximally,reddy2023fractional,wang2024fractional} have presented similar phase diagrams. However, notable discrepancies emerged regarding the bandwidth of the first moir\'e band at a twist angle of $\theta = 3.89^\circ$, with values ranging from 9 to 18 meV. These variations primarily originate from differences in handling vdW corrections, as well as from the selection between plane-wave and localized basis sets. Moreover, lattice relaxation at smaller twist angles significantly influences electronic band structures and topological properties. Due to the substantial number of atoms involved at these small-angle regions, structural relaxation often necessitates advanced machine-learning techniques, and parameters required for continuum models also rely on density functional theory (DFT) calculations~\cite{jia2024moire,mao2024transfer,zhang2024polarization}. Based on that, Xu et al.~\cite{xu2025multiple} and Wang et al.~\cite{wang2025higher} extended the predictions to smaller-angle region, identifying the emergence of more complex phases, including triple, quadruple, and quintuple QSH states. Specifically, at a twist angle of $1.89^\circ$, five distinct bands were each found to carry a Chern number $C=1$ (Fig.~\ref{theory_twist}\textbf{e}). These predictions were further confirmed by nanoribbon calculations, where edge-state analysis clearly demonstrated the realization of a quintuple QSH phase (Fig.~\ref{theory_twist}\textbf{f}). These theoretical advances underscore the rich complexity and diversity of phase diagrams in twisted TMD systems, highlighting their considerable potential for spintronic applications. 

To explore the interaction-driven topological phases, a common approach is to construct effective tight-binding models and incorporate Hubbard-type interactions~\cite{devakul2021magic,xu2024maximally,qiu2023interaction}. Notably, in both twisted-MoTe$_2$ and twisted-WSe$_2$, the top two/three moir\'e valence bands are well separated from the rest of the electronic spectrum. Considering the constraints imposed by the C$_3$ symmetry, the Wannier centers for the first two bands should naturally be located at the MX and XM stacking regions, while the Wannier center for the third band should be placed at the MM region. The MX and XM stacking regions form a moir\'e honeycomb lattice, naturally inheriting a graphene-like lattice geometry. With these three Wannier orbitals, an effective tight-binding Hamiltonian can be expressed as~\cite{devakul2021magic}:
\begin{equation}
H_t = t \sum_{\langle i,j \rangle, \sigma} c_{i \sigma}^{\dagger} c_{j \sigma} + |t_2| \sum_{\langle\langle i,j \rangle\rangle, \sigma} e^{i \phi_\sigma \nu_{ij}} c_{i \sigma}^{\dagger} c_{j \sigma} + \text{h.c.},
\end{equation}
where $c_{i \sigma}^{\dagger}, c_{i \sigma}$ are fermionic creation and annihilation operators at lattice sites i, and $\sigma = \pm$ labels the spin/valley degrees of freedom. The $\nu_{ij} = \pm 1$ indicates the direction of hopping. To incorporate interactions within these flat bands, a simple on-site Hubbard interaction term $H_U = U \sum_i n_{i\uparrow} n_{i\downarrow}$ can be included. Since the interaction strength $U$ is much larger than the hopping amplitudes, the model can be regarded as a strong-coupling limit of the Kane-Mele-Hubbard model~\cite{kane2005quantum,Kane2005z}. By tuning parameters such as the twist angle, displacement field, and interaction strength, the system exhibits a rich landscape of strongly correlated phases, which include the QAH insulator, Mott insulator, as well as ferromagnetic and antiferromagnetic phases~\cite{devakul2021magic}. 

\newpage
\section{Experimental verification of QSH insulator}

As discussed above, QSH insulators are characterized by a gapped 2D bulk band structure and 1D spin-momentum-locked helical edge states. Demonstrating the QSH state requires showing that the 2D bulk state is gapped, confirming the existence of edge states within the 2D bulk gap, and ultimately verifying that the edge states conduct with quantized conductance and exhibit spin-momentum locking.

However, unambiguously detecting the QSH state has been a challenging task due to several reasons. First, the QSH edge conduction is delicate and can be destroyed by time-reversal-breaking scattering events, making quantized conductance observable only in short-channel devices. Second, the QSH state is a 2D topological phase with 1D Dirac cones, making it difficult for techniques like conventional ARPES to directly measure the Dirac cone of the edge states. To overcome these challenges, researchers have employed a suite of complementary techniques. ARPES can resolve the bulk bandgap, while STM/STS provide spatially resolved access to edge-localized density of states. Additionally, microwave impedance microscopy (MIM) enables the distinction between insulating bulk and conductive edge channels. Together with transport measurements, these tools offer a more complete characterization of QSH states~\cite{autes2016novel,tang2017quantum,chen2018large,shi2019imaging,shumiya2022evidence,barber2022microwave,nuckolls2024microscopic,chen2024strong,hu2025high,nuckolls2025spectroscopy}.

\subsection{Spectroscopic and microscopic tools}

ARPES is an essential technique for studying QSH states in 2D topological insulators. By measuring the energy and momentum distribution of photoemitted valence electrons, ARPES provides direct measure of the electronic band structure. Its surface sensitivity makes it well suited for probing atomically thin 2D QSH insulators, enabling direct visualization of key topological features such as band inversion. 

\begin{figure}
\centering
\includegraphics[width=0.9\textwidth]{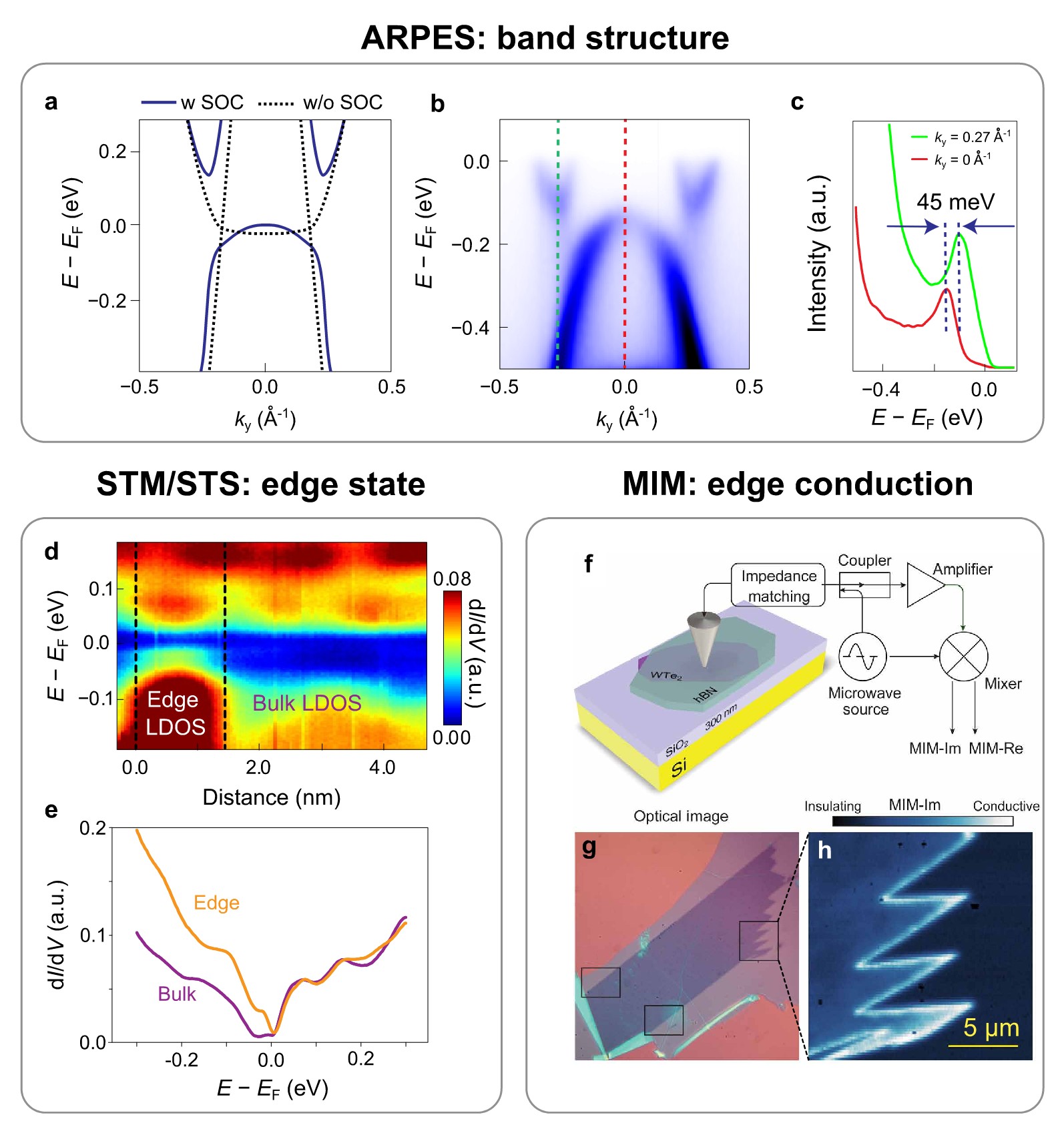}
\caption{\textbf{Spectroscopy and microscopy characterization of a QSH insulator.}
\textbf{a,} Calculated band structure of 1T$^{\prime}$-WTe$_2$ along the $\Gamma$–Y direction. 
\textbf{b,} ARPES data along the $\Gamma$–Y direction taken from epitaxial monolayer 1T$^{\prime}$-WTe$_2$ on bilayer graphene surface. 
\textbf{c,} Energy distribution curves from the momentum positions marked with red and green lines in panel \textbf{b}. The green lines corresponds to the conductance band bottom and the red line corresponds to the valence band top. 
\textbf{d,} $dI/dV$ spectra taken across the step edge of a 1T$^{\prime}$-WTe$_2$ monolayer island.
\textbf{e,} Representative $dI/dV$ spectra taken at the edge (orange) and in the bulk (purple), respectively. 
\textbf{f,} Schematics of the microwave impedance microscopy (MIM) technique. 
\textbf{g,} Optical image of a monolayer WTe$_2$ flake covered with a thin BN protective layer. 
\textbf{h,} Corresponding MIM image of the regions marked in panel \textbf{g}.
Note that Figs.~\textbf{(a-e)} are adapted from Ref.~\cite{tang2017quantum}, Figs.~\textbf{(f-h)} from Ref.~\cite{shi2019imaging}.
}
\label{Sepctroscopy_QSH}
\end{figure}

Tang et al.~\cite{tang2017quantum} reported ARPES measurements of monolayer 1T$^{\prime}$-WTe$_2$ in 2017, with the sample grown by molecular beam epitaxy on a bilayer graphene substrate. The measured electronic structure agrees well with theoretical calculations (Figs.~\ref{Sepctroscopy_QSH}\textbf{a-c}), exhibiting clear band inversion and a well-defined bulk band gap. Specifically, the conduction and valence bands are well separated, with an estimated gap of $\sim 45 \pm 20$~meV (Fig.~\ref{Sepctroscopy_QSH}\textbf{c}). Autès et al.~\cite{autes2016novel} investigated the topological edge states in the quasi-1D topological insulator $\beta$-Bi$_4$I$_4$ using ARPES. Their measurements revealed a Dirac cone at the $\textbf{M}$ point of the surface Brillouin zone, confirming the material’s strong topological insulator phase with well-defined $\mathcal{Z}_2$ invariants. The surface states are also found to be strongly anisotropic, consistent with theoretical predictions.

STM/STS and MIM are particularly useful for distinguishing between bulk and edge states and conduction~\cite{tang2017quantum, ma2015unexpected, shi2019imaging}. In the same study that reported the ARPES measurements~\cite{tang2017quantum}, Tang et al. also performed spatially resolved differential conductance ($dI/dV$) measurements across the edge of a monolayer 1T$^{\prime}$-WTe$_2$ island (Figs.~\ref{Sepctroscopy_QSH}\textbf{d-e}). The $dI/dV$ spectra in the bulk exhibit a nearly flat signal close to zero at the Fermi level, consistent with an insulating gap, and the peak positions match the band edges identified by ARPES. In contrast, the spectra at the edge exhibit a distinct V-shaped profile, with finite spectral weight inside the bulk gap and a spatial confinement of approximately 1.5~nm, indicating the presence of edge states.

The STM/STS technique has been further employed to investigate the nature of the topological gap in monolayer WTe$_2$~\cite{jia2022evidence,que2024gate}. Jia et al.~\cite{jia2022evidence} reported an insulating gap of 47--91~meV that remains pinned at zero bias regardless of Fermi level tuning and vanishes under strong doping. These observations are inconsistent with a conventional band insulator and instead point to a correlation-driven excitonic insulator phase, in which spontaneous electron-hole pairing forms a many-body ground state. Subsequent STS studies by Que et al.~\cite{que2024gate} further support the existence of a topological excitonic insulator. An ambipolar quantum phase transition was observed, where the gap collapses abruptly under either electron or hole doping, transforming WTe$_2$ into an $n$-type or $p$-type semimetal. Notably, metallic edge states persist across the transition, consistent with the robust topological character of WTe$_2$. These findings establish monolayer WTe$_2$ as a tunable platform for exploring correlated topological phases, including exciton-mediated superconductivity, and highlight the intricate interplay between strong electronic interactions and topology in 2D systems.

Similar ARPES and STM/STS studies have also been conducted on other QSH systems, such as 1T$^{\prime}$ phase WSe$_2$, Bi$_4$Br$_4$, etc. Chen et al.~\cite{chen2018large} observe a 129 meV band gap and an in-gap edge state near the layer boundary of a quasi-freestanding monolayer 1T$^{\prime}$ phase WSe$_2$ grown on bilayer graphene, in contrast to the semiconducting 1H phase WSe$_2$ with a trivial bandgap ($\sim2$ eV). Shumiya et al.~\cite{shumiya2022evidence} use STM to provide evidence for a room-temperature QSH edge state on the surface of the higher-order topological insulator Bi$_4$Br$_4$. The atomically resolved lattice exhibits a large insulating gap of over 200 meV, and an atomically sharp monolayer step edge hosts an in-gap gapless state, consistent with topological bulk–boundary correspondence. These large-gap QSH insulators offer promising platforms for exploring high-temperature transport quantization and advancing topological electronics.

The MIM scanning probe technique measures tip–sample
admittance (inverse of impedance) in a non-contact geometry at microwave frequencies (Fig.~\ref{Sepctroscopy_QSH}\textbf{f}) ~\cite{ma2015unexpected, shi2019imaging,sun2022evidence,barber2022microwave}, and reveals the contrast between bulk and edge conductance. Shi et al.~\cite{shi2019imaging} first apply the MIM technique to monolayer WTe$_2$ devices exfoliated on SiO$_2$ substrate with a thin boron nitride (BN) as a protective layer for avoiding degradation (Fig.~\ref{Sepctroscopy_QSH}\textbf{g}). The observation of conduction localized at the sample edges (Fig.~\ref{Sepctroscopy_QSH}\textbf{h}) is consistent with the QSH effect. This technique has also proven valuable as a diagnostic tool for identifying internal structural features, such as cracks in WTe$_2$, which create unintended conduction channels not detectable by global transport measurements~\cite{shi2019imaging}. Moreover, it has been extended to investigate other systems, including correlated insulating phases~\cite{huang2021correlated} and fractional QAH states~\cite{ji2024local}.

Mid-infrared absorption micro-spectroscopy and pump–probe micro-spectroscopy are also powerful optical techniques used to study the QSH insulators like Bi$_4$Br$_4$~\cite{han2023optical}. Mid-infrared absorption micro-spectroscopy was able to distinguish the bulk and boundary states in Bi$_4$Br$_4$ by revealing strong absorption from the gapless edge states, while the bulk shows suppressed absorption due to its insulating gap. Pump–probe micro-spectroscopy, on the other hand, measures carrier dynamics, showing that the bulk exhibits fast relaxation (picoseconds), whereas the boundary states display an ultralong carrier lifetime (up to 1.5 nanoseconds), perhaps due to its topological nature. This technique reveals the potential of QSH insulators in infrared and THz optoelectronic applications.

\subsection{Transport and quantized edge conductance}

Compelling evidence for nontrivial topology is expected to be obtained from transport measurements. Within the QSH gap, we anticipate observing residual edge conductance and an enhanced nonlocal response, as well as quantized edge conductance in the ballistic transport regime. As representative examples, we highlight transport signatures observed in monolayer WTe$_2$ and TaIrTe$_4$ QSH systems.

(1) Residual edge conductance within the bandgap. In a QSH insulator, tuning the chemical potential into the bandgap suppresses bulk conduction, allowing edge modes to dominate transport. This results in a nonzero residual conductance within the gap, as demonstrated in monolayer WTe$_2$ and TaIrTe$_4$ devices. Fei et al.~\cite{fei2017edge} first reported residual edge conduction in hexagonal BN-encapsulated monolayer WTe$_2$, in contrast to bilayer WTe$_2$, which is topologically trivial and exhibits nearly zero conductance inside the gap. Tang et al.~\cite{Tang2024DualQSH} observed similar behavior in TaIrTe$_4$, where the residual conductance remains nearly temperature-independent below 30~K. By subtracting the residual contribution, the remaining temperature-dependent conductance can be analyzed using Arrhenius fitting to estimate the bulk gap, yielding an activation energy of 20–30~meV in TaIrTe$_4$, consistent with first-principles calculations~\cite{liu2017van}.

(2) Enhancement of the nonlocal response inside the gap. Nonlocal transport measurements are a powerful probe of edge conduction in QSH systems~\cite{fei2017edge,sun2022evidence,Tang2024DualQSH}. When bulk conduction dominates, current flows primarily through the interior and remains largely confined between the current injection and collection contacts, resulting in a weak nonlocal signal. In contrast, when the bulk becomes insulating and edge states dominate, current is forced to propagate along the sample boundaries, leading to a significantly enhanced nonlocal response. To quantitatively assess this behavior, simultaneous measurements of local and nonlocal voltages are typically performed, and their ratio is evaluated. For example, in TaIrTe$_4$, Tang et al.~\cite{Tang2024DualQSH} observed a pronounced enhancement of this ratio within the QSH gap, consistent with dominant edge conduction along the sample boundary.

(3) Quantized edge conductance in the ballistic transport regime. A hallmark of the QSH state is the presence of a single conducting channel per edge when carrying directional current, leading to a quantized conductance of $e^2/h$ per edge ($G_0$). This quantization provides unambiguous evidence of the QSH effect, distinguishing it from conventional edge transport, where both spin up and down modes propagate in the same direction, yielding a conductance of $2Ne^2/h$ per edge ($N$ is an integer). In contrast, QSH insulators host spin-momentum locked edge channels, which suppress backscattering and enable more robust edge transport.

\begin{figure}
\centering
\includegraphics[width=\textwidth]{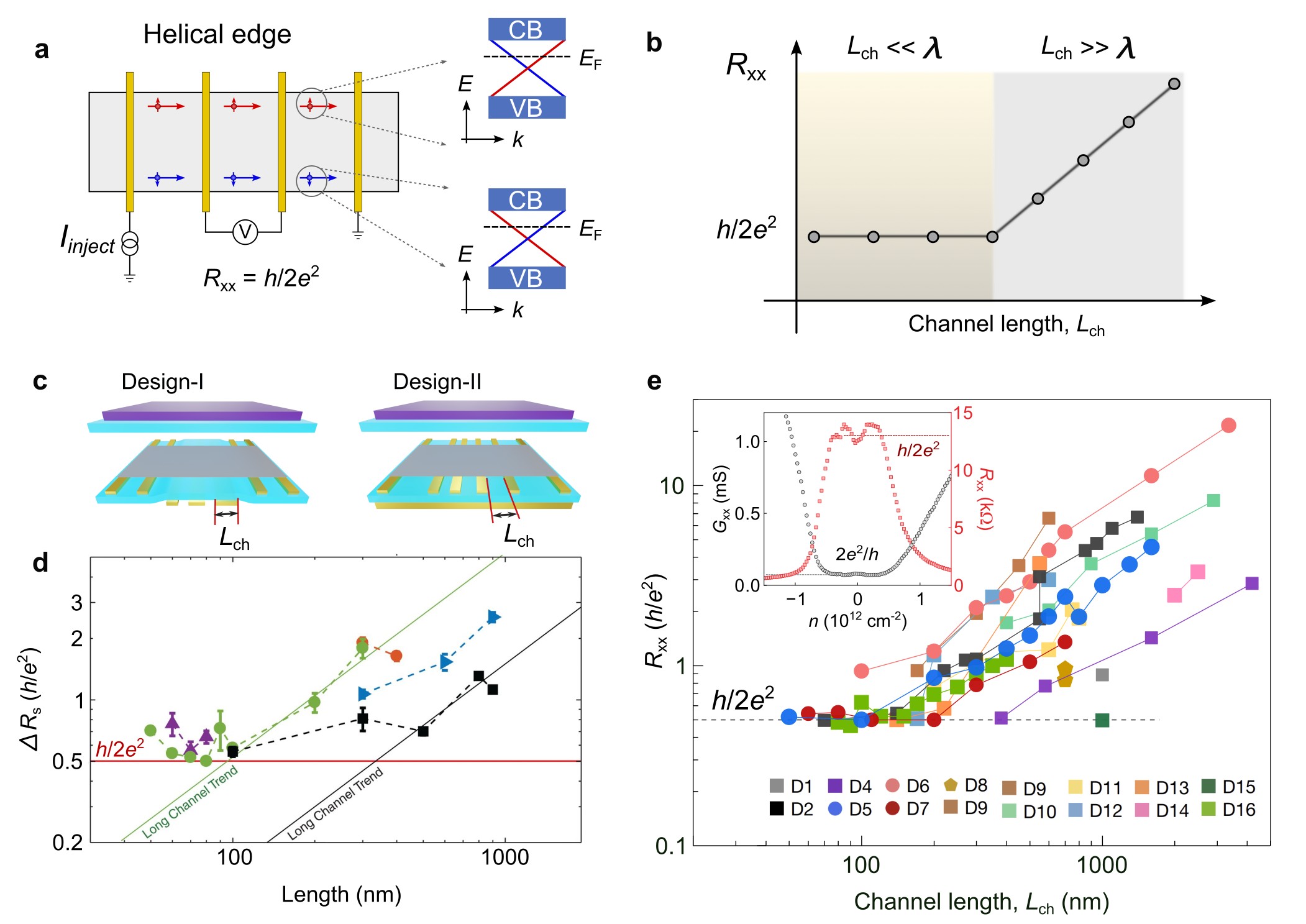}
\caption{\textbf{Quantization of the QSH edge state.}
\textbf{a,} The quantized edge conduction of $2e^2/h$ ($e^2/h$ per edge) is a hallmark of QSH.
\textbf{b,} A schematic illustration of the edge conduction versus channel length. In the ballistic transport regime ($L_{ch} \ll \lambda$), the edge conduction quantize at $2e^2/h$, while increasing the channel length introduces scattering, leading to a gradual deviation from quantization.
\textbf{c,} Schematic illustrations of two kinds of short-channel device: a gate-defined channel with $L_{ch}$ determined by the gate width (Design-I) and a contact-defined channel with $L_{ch}$ determined by the distance between two adjacent contacts (Design-II).
\textbf{d,} Channel-length dependence of the undoped-channel resistance for the WTe$_2$ devices. The edge conductance approach $2e^2/h$ for $L_{ch} \sim 100$ nm.
\textbf{e,} Channel-length dependence of the plateau resistance at the CNP for the TaIrTe$_4$ devices. The edge conductance approach $2e^2/h$ for $L_{ch} \sim 200$ nm.
Note that Figs.~\textbf{(a-c, e)} are adapted from Ref.~\cite{Tang2024DualQSH}, Fig.~\textbf{(d)} from Ref.~\cite{wu2018observation}.
}
\label{channel_length}
\end{figure}

However, it has been found that QSH edge states are not perfectly robust, and careful device engineering is required to reach the ballistic transport regime ($L_{\mathrm{ch}} \ll \lambda$, Figs.~\ref{channel_length}\textbf{a,b}), where $L_{\mathrm{ch}}$ is the channel length and $\lambda$ is the edge mean free path. Structural and chemical instabilities of monolayers, variations in crystal quality, and challenges in achieving high-quality ohmic contacts further complicate the realization of ballistic devices. Wu et al.~\cite{wu2018observation} addressed the challenge of achieving high-quality ohmic contacts in monolayer WTe$_2$ by implementing a gate-defined short-channel design (Fig.~\ref{channel_length}\textbf{c}, Design-I). In this architecture, a global top gate was used to heavily dope the entire sample, while a local bottom gate was applied to locally neutralize the charge density, effectively defining the channel length by the width of the bottom gate. This approach enabled the clear observation of quantized conductance of $e^2/h$ per edge for $L_{\mathrm{ch}} < 100$~nm in monolayer WTe$_2$, which notably persists up to 100~K (Fig.~\ref{channel_length}\textbf{d}).

Tang et al.~\cite{Tang2024DualQSH} implemented both gate-defined (Design-I) and contact-defined (Design-II) short-channel architectures in monolayer TaIrTe$_4$ devices (Fig.~\ref{channel_length}\textbf{c}). In Design-II, the channel length was defined by the spacing between adjacent metal contacts. Both designs exhibited quantized edge conductance of approximately $2e^2/h$ in the ballistic regime for $L_{\mathrm{ch}} < 200$~nm, confirming the QSH effect in TaIrTe$_4$ (Fig.~\ref{channel_length}\textbf{e}).

To further verify the spin-momentum locking of QSH edge states, nonlocal spin valve measurements were performed using ferromagnetic contacts (e.g., Fe$_3$GeTe$_2$) to inject and detect spin-polarized currents in TaIrTe$_4$~\cite{Tang2024DualQSH}. Due to spin-momentum locking, the edge current is carried by spin-polarized electrons. When an external magnetic field is swept to reverse the magnetization of the ferromagnetic electrodes, the conductance depends on the relative alignment between the contact spin polarization and the spin of the edge current. This leads to a hysteretic response in the nonlocal resistance as a function of magnetic field, as observed in~\cite{Tang2024DualQSH}. Furthermore, reversing the current direction reverses the spin polarization carried by the edge current, resulting in a reversal of the hysteresis.

\subsection{QSH insulator under magnetic field}

It is commonly believed that QSH helical edge conduction is protected by time-reversal symmetry. Applying a magnetic field is expected to open a gap at the edge Dirac point, thereby suppressing edge conduction. However, in real materials, the situation is more complex and depends on factors such as the edge spin orientation and the energy position of the edge Dirac point, among others~\cite{fei2017edge,wu2018observation,zhao2021determination,Tang2024DualQSH}. Furthermore, the QSH state may evolve into QAH state upon applying magnetic field~\cite{zhao2024realization}. Below, we discuss three cases illustrating how magnetic fields modify QSH edge conduction in the materials covered in the following sections.

\begin{figure}
\centering
\includegraphics[width=\textwidth]{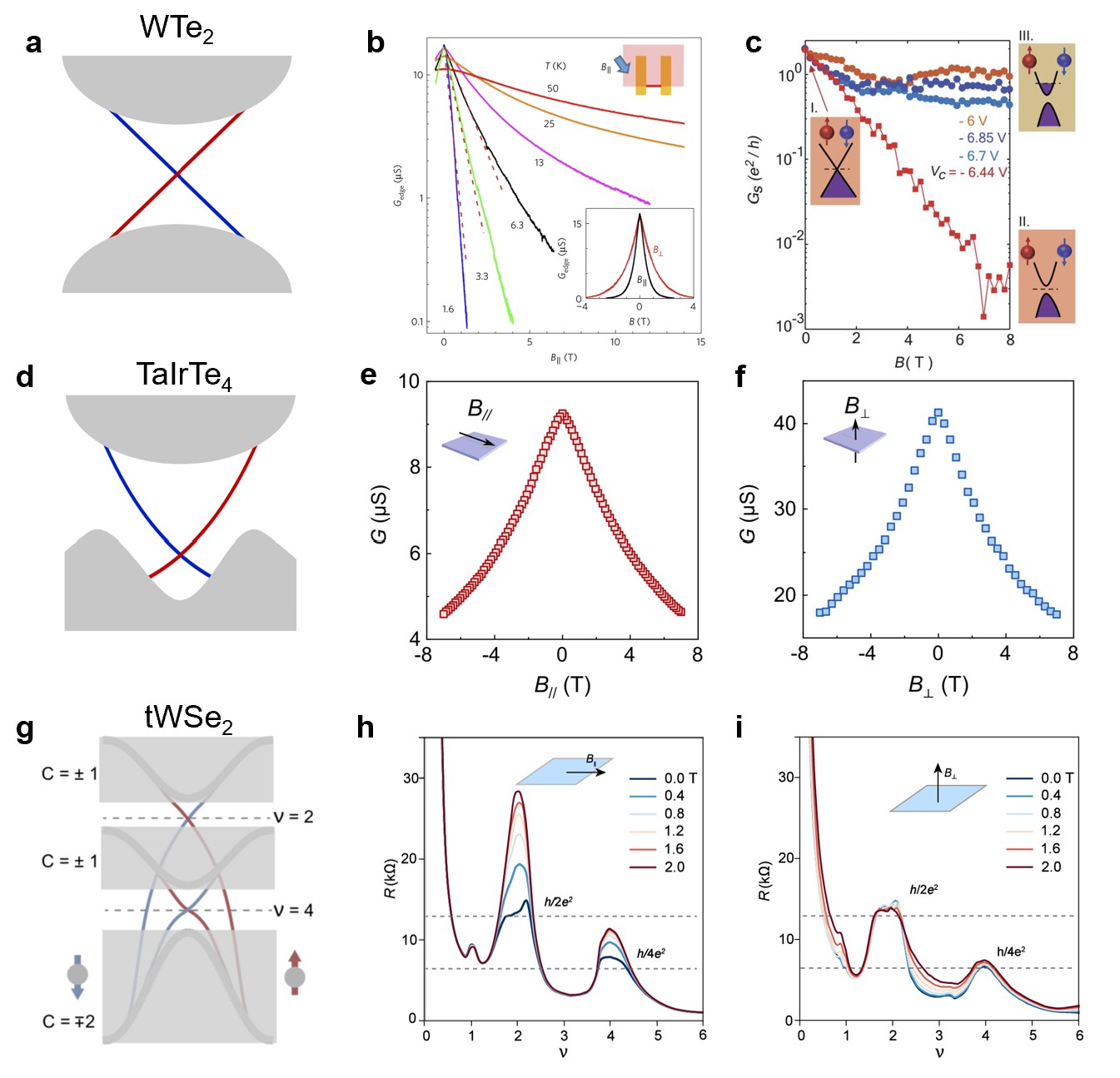}
\caption{\textbf{QSH edge state under magnetic field.}
\textbf{a,} The schematic of the crossing point residues within the band gap.
\textbf{b-c,} The edge conductance of WTe$_2$ under the parallel magnetic field ($B_{\parallel}$) for various temperatures and under the parallel magnetic field ($B_{\perp}$).
\textbf{d,} The schematic of the crossing point residues within the valence band.
\textbf{e-f,} The edge conductance of TaIrTe$_2$ under $B_{\parallel}$ and $B_{\perp}$.
\textbf{g,} The schematic of QSH effect with $S_z$ conservation.
\textbf{h-i,} The edge conductance versus filling factor $\nu$ of twist WSe$_2$ device under $B_{\parallel}$ and $B_{\perp}$.
Note that Fig.~\textbf{(b)} is adapted from Ref.~\cite{fei2017edge}, Fig.~\textbf{(c)} from Ref.~\cite{wu2018observation}, Figs.~\textbf{(e-f)} from Ref.~\cite{Tang2024DualQSH}, Figs.~\textbf{(g-i)} from Ref.~\cite{kang2024double}.
}
\label{edge_under_Bfield}
\end{figure}

In monolayer WTe$_2$, both Fei \textit{et al.}~\cite{fei2017edge} and Wu \textit{et al.}~\cite{wu2018observation} reported an exponential decrease in edge conduction within the QSH regime under an applied out-of-plane magnetic field (Figs.~\ref{edge_under_Bfield}\textbf{b-c}). This behavior is consistent with a Zeeman-induced gap opening in the edge states, with effective $g$-factors of approximately $g = 7.5$ and $g = 4.8$, respectively (Fig.~\ref{edge_under_Bfield}\textbf{a}). Additionally, Zhao \textit{et al.}~\cite{zhao2021determination} applied a vector magnetic field to monolayer WTe$_2$ to probe the spin texture of the edge modes. They observed that edge conductance is most strongly suppressed by the field component perpendicular to a specific axis, indicating that the edge spins align along this direction. Remarkably, this axis remains fixed relative to the crystal structure—lying at ($40 \pm 2$)$^\circ$ to the layer normal within the mirror plane—regardless of edge orientation, chemical potential, and other variables. This observation points to a momentum-independent spin axis in the edge modes~\cite{garcia2020canted}.

In monolayer TaIrTe$_4$, Tang \textit{et al.}~\cite{Tang2024DualQSH} reported a weak suppression of edge conduction under both in-plane and out-of-plane magnetic fields (Figs.~\ref{edge_under_Bfield}\textbf{e,f}), with an effective $g$-factor of approximately $g = 0.6 \sim 0.7$. A simple explanation, supported by DFT calculations, is that the edge Dirac point lies within the bulk valence band (Fig.~\ref{edge_under_Bfield}\textbf{d}). In this case, when a magnetic field induces a gap, the gap is embedded within or very close to the bulk valence band and thus not directly observable. Nevertheless, applying a magnetic field allows backscattering between opposite spins, leading to a suppression of edge conduction, albeit in a relatively weak manner.

In moir\'e TMD systems, QSH edge states are thought to be protected by Ising spin conservation ($S_z$ conservation) (Fig.~\ref{edge_under_Bfield}\textbf{g}). Kang \textit{et al.}~\cite{kang2024double} observed edge resistance plateaus at $\nu = 2$ and 4 that are insensitive to out-of-plane magnetic fields but increase under in-plane fields (Figs.~\ref{edge_under_Bfield}\textbf{h-i}). The out-of-plane magnetic field $B_{\perp}$, aligned with the spin quantization axis, does not affect the helical edge states protected by $S_z$ conservation. In contrast, the in-plane magnetic field induces spin mixing and backscattering, leading to the localization of edge states and a transport gap~\cite{zhao2021determination}.

\section{Recent Representative QSH insulator Materials}

In this section, we discuss the recent discovery of three representative QSH material systems: 1T$^\prime$-WTe$_2$, TaIrTe$_4$, and twisted TMD moiré systems. These systems exhibit intriguing quantum phases intertwined with the QSH phase. 

In monolayer WTe$_2$, the topological gap is believed to have a Coulomb origin, with one possibility being exciton condensation~\cite{jia2022evidence,sun2022evidence}. Unexpected quantum oscillations and Landau quantizations have been observed within the gap, suggesting the possible existence of a charge-neutral Fermi surface~\cite{wang2021landau}. Furthermore, electron doping at densities below $10^{13}$ cm$^{-2}$ induces a superconducting phase~\cite{fatemi2018electrically,sajadi2018gate,song2024unconventional}. Stacking engineering in WTe$_2$ has also revealed fascinating phenomena: the spin splitting can be controlled by parallel or antiparallel stacking of bilayers, as demonstrated by ARPES measurements~\cite{zhang2023symmetry}. Due to the anisotropic lattice of monolayer WTe$_2$, twisted bilayer WTe$_2$ forms a moiré lattice that acts as a 2D array of 1D conduction channels, exhibiting 1D Luttinger liquid behavior~\cite{wang2022one,yu2023evidence}. Monolayer TaIrTe$_4$ is a recently realized QSH insulator~\cite{liu2017van, guo2020quantum, Tang2024DualQSH}. Uniquely, in addition to the single-particle QSH state at charge neutrality, a correlated QSH phase emerges at finite doping, driven by the correlation effects near van Hove singularities (VHSs) and possibly a charge density wave (CDW) phase~\cite{Tang2024DualQSH}. In moiré platforms, double and triple QSH states have been observed in twisted bilayer MoTe$_2$, WSe$_2$, and hetero-bilayer systems~\cite{zhao2024realization,kang2024evidence,kang2024double}. Remarkably, the QSH bands can transform into Chern bands, leading to topological phase transitions by applying an out-of-plane magnetic field~\cite{zhao2024realization}.

\subsection{WTe$_2$: QSH, excitonic insulating gap and superconductivity} \label{subsec:WTe2}

The binary 1T$^\prime$-WX$_2$ family was proposed in 2014~\cite{Qian2014Quantum}, and since then, significant efforts have been made to explore this family. While other members of the 1T$^\prime$ phase, such as 1T$^\prime$-WSe$_2$, have also been investigated~\cite{ugeda2018observation,chen2018large}, a particularly successful case is monolayer WTe$_2$, where robust band inversion, driven by a distorted lattice and strong SOC, underpins its non-trivial topology (Fig.~\ref{Sepctroscopy_QSH}\textbf{a}). Interestingly, initial theoretical calculations predicted a negative gap, which, in principle, would prevent the experimental detection of edge states~\cite{Qian2014Quantum}. However, experiments using ARPES, STM, and transport measurements~\cite{tang2017quantum,fei2017edge,wang2021landau,jia2022evidence,sun2022evidence} have consistently observed a large gap on the order of tens of meV.

\begin{figure}
\centering
\includegraphics[width=\textwidth]{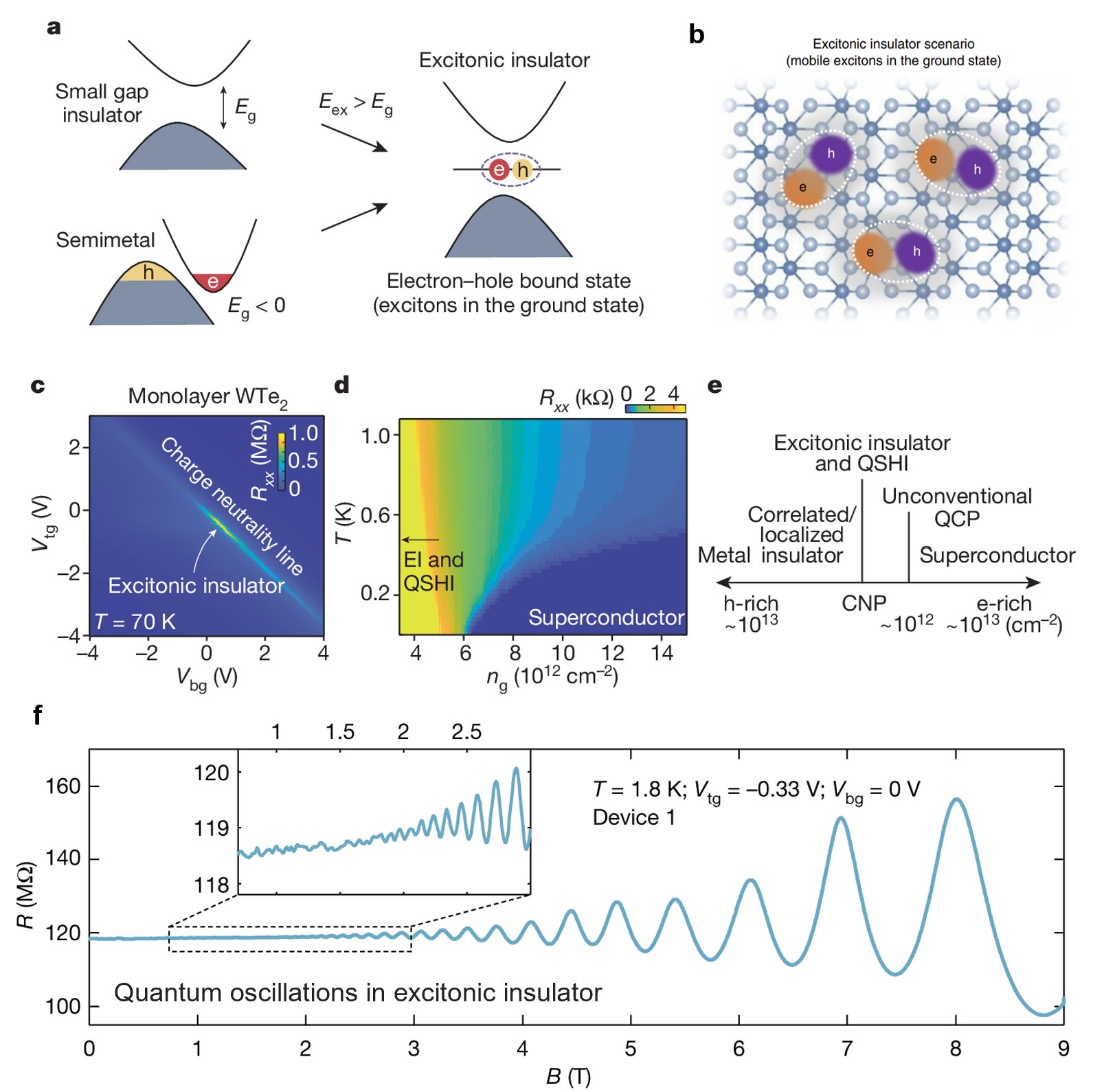}
\caption{\textbf{Excitonic insulator and superconductivity in monolayer WTe$_2$.}
\textbf{a,} Schematic illustration of exciton formation in small-gap insulators or semimetals.
\textbf{b,} Illustration of the excitonic insulator phase in monolayer WTe$_2$. 
\textbf{c,} Dual-gate resistance map of monolayer WTe$_2$, highlighting the excitonic insulating state at charge neutrality. 
\textbf{d,} Resistance phase diagram of monolayer WTe$_2$ under electron doping, showing the emergence of superconductivity adjacent to the excitonic insulator and QSH.
\textbf{e,} Phase diagram of monolayer WTe$_2$ covering both electron- and hole-doped regimes.
\textbf{f,} Low-temperature magnetoresistance curve displaying quantum oscillations in the insulating regime of a monolayer WTe$_2$ device. 
Note that Figs.~\textbf{(a,c-e)} are adapted from Ref.~\cite{wu2024charge}, Fig.~\textbf{(b)} from Ref.~\cite{jia2022evidence}, Fig.~\textbf{(f)} from Ref.~\cite{wang2021landau}.
}
\label{excitonic insulator}
\end{figure}

The nature of this gap has been a topic of significant interest in the study of WTe$_2$. While adjustments to DFT calculations can reproduce a gap at the single-particle level, experimental evidence suggests that the gap is interaction-induced rather than a single-particle effect. A particularly intriguing possibility is that WTe$_2$ is an excitonic insulator (Fig.~\ref{excitonic insulator}\textbf{b}). Excitons are quasiparticles formed when an electron in the conduction band binds to a hole in the valence band due to Coulomb attraction~\cite{pereira2022topological,wu2024charge}. An excitonic insulator is a phase of matter that emerges when bound excitons condense, behaving as charge-neutral bosons at low densities. In the single-particle picture of a conventional insulator or semiconductor, a finite energy gap $E_g$ separates the fully occupied valence band from the conduction band. However, when the band gap is small and the exciton binding energy $E_{ex}$ exceeds $E_g$, Coulomb interactions drive the formation of bound electron-hole pairs (excitons), as illustrated in Figs.~\ref{excitonic insulator}\textbf{a-b}. A similar scenario could also occur in low carrier-density semimetals~\cite{kogar2017signatures,lu2017zero}.

Experimentally, Jia \textit{et al.}~\cite{jia2022evidence} provided evidence supporting monolayer WTe$_2$ as an excitonic insulator (Fig.~\ref{excitonic insulator}\textbf{c}). First, they observed an insulating temperature dependence in $R_{xx}$, with $R_{xx}$ diverging at low temperatures, indicating a large gap, while $R_{xy}$ remained finite without divergence or sign reversal near the gap—contrary to expectations for a conventional band insulator. This behavior could result from excitonic pairing between electrons and holes, whose opposing Hall contributions nearly cancel. Second, the excitonic insulating behavior was further supported by gate-tuned tunneling spectroscopy. These measurements revealed an insulating gap pinned at zero bias, which collapsed abruptly under strong electron or hole doping—indicating a correlation-driven gap at zero or very low dopings. This was also discussed previously in Section~\textbf{3.1}. In parallel, Sun \textit{et al.}~\cite{sun2022evidence} investigated the chemical potential as a function of doping and observed a sharp step ($\sim$40~meV) at the CNP, emerging below 100~K—contradicting the gradual transition expected in conventional band insulators. This behavior, together with a V-shaped conductivity profile observed above 100~K, indicates strong electron–hole correlations and the formation of equilibrium excitons.

Beyond the topological excitonic insulator phase, monolayer WTe$_2$ also exhibits superconductivity at sub-kelvin temperatures when a moderate electron density is introduced via electrostatic gating (Fig.~\ref{excitonic insulator}\textbf{d})~\cite{fatemi2018electrically,sajadi2018gate,song2024unconventional}. Fatemi et al.~\cite{fatemi2018electrically} and Sajadi et al.~\cite{sajadi2018gate} observed a gate-tuned transition from a QSH insulator to a superconducting phase. While the superconductivity’s topological nature remains unconfirmed, the continuous transition from a topological insulator to an intrinsic superconductor opens possibilities for realizing topological superconductivity~\cite{qiu2021recent,hsu2017topological}. These findings suggest potential device applications where local gating enables the coexistence of topological and superconducting phases in the same material, offering a pathway to explore non-Abelian topological states and develop gateable superconducting circuits within a single material system.

The diverse quantum phases in monolayer WTe$_2$ make it a compelling platform for investigating the interplay between excitonic insulators, topology, and superconductivity (Fig.~\ref{excitonic insulator}\textbf{e}). Despite significant progress, a complete understanding of this system is still ongoing. For instance, recent studies have reported unexpected quantum oscillations within the energy gap of monolayer WTe$_2$ (Fig.~\ref{excitonic insulator}\textbf{f}), suggesting the presence of highly mobile, charge-neutral fermionic excitons~\cite{wang2021landau,wu2024detection}. The origin of these oscillations remains under debate. Proposed other explanations include thermally activated carriers within the insulating gap or the coupling of monolayer WTe$_2$ to nearby graphite gates, where it acts as a sensor for Landau levels in graphite~\cite{lee2021quantum,he2021quantum,zhu2021quantum}. However, recent thermoelectric quantum oscillation measurements in the hole-doped regime provide evidence that these mobile carriers originate intrinsically from WTe$_2$, with a Landau level-like energy spectrum emerging at the chemical potential of the insulator under applied magnetic fields~\cite{wu2024detection,tang2024sign}.

Twist-controlled symmetry provides a powerful means of tuning the electronic properties of 2D materials. Zhang et al.~\cite{zhang2023symmetry} fabricated twisted bilayer WTe$_2$ with controlled twist angles to investigate symmetry-dependent electronic responses. In bilayers twisted by 180$^\circ$, inversion symmetry is broken, resulting in a strong second-harmonic generation (SHG) signal. In contrast, the 0$^\circ$ configuration preserves inversion symmetry, suppressing the SHG response. ARPES measurements further revealed substantial spin splitting ($\sim$100~meV) in the 180$^\circ$ twisted bilayers, confirming that symmetry control via twist angle can significantly modify the electronic structure.

In a separate study, Wang et al.~\cite{wang2022one} experimentally realized a 2D array of 1D Luttinger liquids in twisted bilayer WTe$_2$ that form an anisotropic moiré superlattice. At a twist angle near 5$^\circ$, a highly ordered array of parallel 1D electronic channels emerged, with nanoscale spacing between adjacent wires. Transport measurements revealed extreme anisotropy in resistance (up to $\sim$1000$\times$) between two orthogonal in-plane directions, indicating the formation of a strongly anisotropic electronic phase. The conductance along the wires exhibited power-law scaling, consistent with Luttinger liquid behavior. Furthermore, tuning the doping level from hole-doped region to electron-doped region will induce a transition from an anisotropic Luttinger liquid regime to an isotropic electronic state. This work establishes a new platform for investigating coupled-wire models and correlated 1D electron physics in a moiré-engineered setting.

\subsection{TaIrTe$_4$: QSH, van Hove singularities and charge density wave}

Shortly after the prediction of the 1T$^\prime$-MX$_2$ family, Liu et al. proposed a family of layered ternary transition metal chalcogenides, MM$^\prime$Te$_4$ (M = Nb, Ta; M$^\prime$ = Ir, Rh)~\cite{liu2017van,guo2020quantum}. These compounds are Weyl semimetals in their 3D bulk form and QSH insulators with sizable band gaps in their monolayer form. In 2019, STM was first employed to detect the edge density of states at step edges in TaIrTe$_4$~\cite{dong2019observation}, providing supporting evidence for QSH state. Additionally, superconductivity was observed on the surface of TaIrTe$_4$ using STM/STS~\cite{cai2019observation,xing2020surface,tang2021strong}. However, it was not until very recently that monolayers were successfully isolated from the vdW bulk crystals and fabricated into electronic devices for transport studies.

\begin{figure}
\centering
\includegraphics[width=\textwidth]{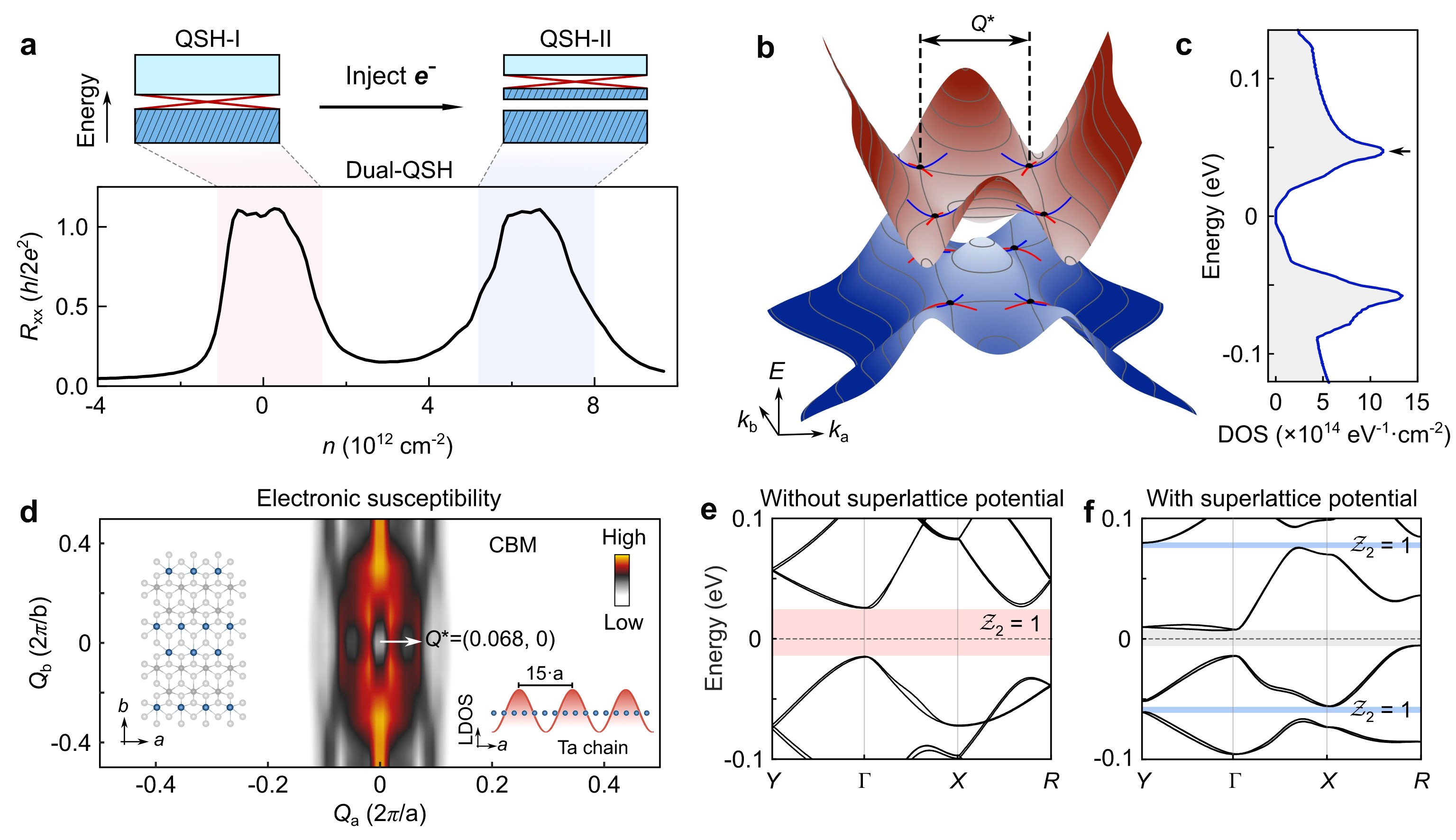}
\caption{\textbf{Dual QSH phase by density-tuned correlations in monolayer TaIrTe$_4$.}
\textbf{a,} The dual QSH effect: Resistance versus carrier density at T=1.7 K (bottom) and a schematic of the dual QSH (top). Two resistance plateaus are developed at the CNP and $n = 6.5 \times 10^{12}$ cm$^{-2}$, corresponding to the QSH-I and QSH-II states, respectively.
\textbf{b,}  The 3D band structure of monolayer TaIrTe$_4$ with VHSs denoted by black dots.
\textbf{c,} The calculated density of states as a function of energy. The black arrow marks the energy level of VHS with a large DOS. 
\textbf{d,} Calculated electronic susceptibility of the conduction band based on the first-principles band structure. The white arrow highlights a local maximum of the electronic susceptibility at a wavevector \textbf{Q}*. \textbf{Q}* connects two neighbouring VHSs as marked in panel \textbf{b}, which corresponds to a superlattice of about 15 unit cells (right inset). 
\textbf{e-f,} The folded band structure with a superlattice periodicity of 15 unit cells, without (\textbf{e}) and with (\textbf{f}) superlattice potential. 
Note that Figs.~\textbf{(a-f)} are adapted from Ref.~\cite{Tang2024DualQSH}.
}
\label{Dual_QSH}
\end{figure}

Systematic transport studies have demonstrated quantized edge conductance consistent with QSH helical states within the single-particle gap (Fig.~\ref{channel_length}). Theoretical analysis and DFT calculations reveal that two Dirac cones at the $\Lambda$ points, gapped by SOC, give rise to the QSH state at the charge neutrality point (CNP). Remarkably, upon finite electron doping, a new insulating gap emerges, within which quantized edge conduction reappears (Fig.~\ref{Dual_QSH}\textbf{a}). This second QSH state is highly unconventional and originates from the interplay between single-particle band topology and VHS-induced electron correlations~\cite{Tang2024DualQSH}. Thus, monolayer TaIrTe$_4$ hosts two distinct QSH phases: one at CNP and another within a correlated insulating state induced by electron doping.

The second QSH state (QSH-II) arises from electron correlations, likely driven by a CDW instability. The low-energy band structure of monolayer TaIrTe$_4$ features three Fermi pockets that expand and merge into saddle points upon doping, forming VHSs with enhanced density of states (Figs.~\ref{Dual_QSH}\textbf{b-c}). Enhanced electronic susceptibility at the VHSs may facilitate CDW formation, with a nesting wavevector $\mathbf{Q}$* (Fig.~\ref{Dual_QSH}\textbf{d}). The resulting CDW superlattice induces band folding and opens a gap (Figs.~\ref{Dual_QSH}\textbf{e-f}). Topological invariant calculations confirm that the gap remains topologically nontrivial. Intuitively, this arises because the nesting wavevector $\mathbf{Q}$* lies within the band inversion regime, allowing the correlated gap to inherit the nontrivial topology of the original bands.

The discovery of a dual-QSH insulator in monolayer TaIrTe$_4$ establishes a new platform for engineering topological flat minibands and studying the interplay between topology and charge-order superlattices. The flat bands associated with the CDW superlattice, characterized by nanometer-scale periodicity, bear a strong resemblance to moiré flat bands. This system provides a promising route to explore time-reversal-invariant fractional topological phases and emergent electromagnetic responses, especially at fractional fillings and ultralow temperatures in high-quality samples~\cite{cai2023signatures,zeng2023thermodynamic,park2023observation,xu2023observation,kang2024evidence}.

Inspired by these findings, researchers predicted a similar dual-QSH insulator state in NbIrTe$_4$, an isoelectronic counterpart of TaIrTe$_4$. First-principles calculations~\cite{liu2024prediction} reveal that NbIrTe$_4$ exhibits not only a single-partile $\mathcal{Z}_2$ band gap at charge neutrality but also a VHS-induced correlated $\mathcal{Z}_2$ band gap under weak doping conditions. The VHS-induced band gap in NbIrTe$_4$ is dominated by Nb-4$d$ orbitals, suggesting strong electronic correlations. In fact, the MM$^\prime$Te$_4$ family compounds are all promising candidates for further exploration of correlated topological phenomena, with potential implications for observing fractional QSH effects and helical quantum spin liquids. Moreover, the emergent correlated insulating phase may spontaneously break various symmetries. When combined with the large Berry curvature of the topological bands, such symmetry breaking can give rise to pronounced linear and nonlinear electromagnetic responses at low frequencies, originating from quantum geometric effects~\cite{sodemann2015quantum, ma2019observation, kang2019nonlinear}.

\subsection{QSH insulator and Chern insulator in moir\'e systems}

\begin{figure}
\centering
\includegraphics[width=\textwidth]{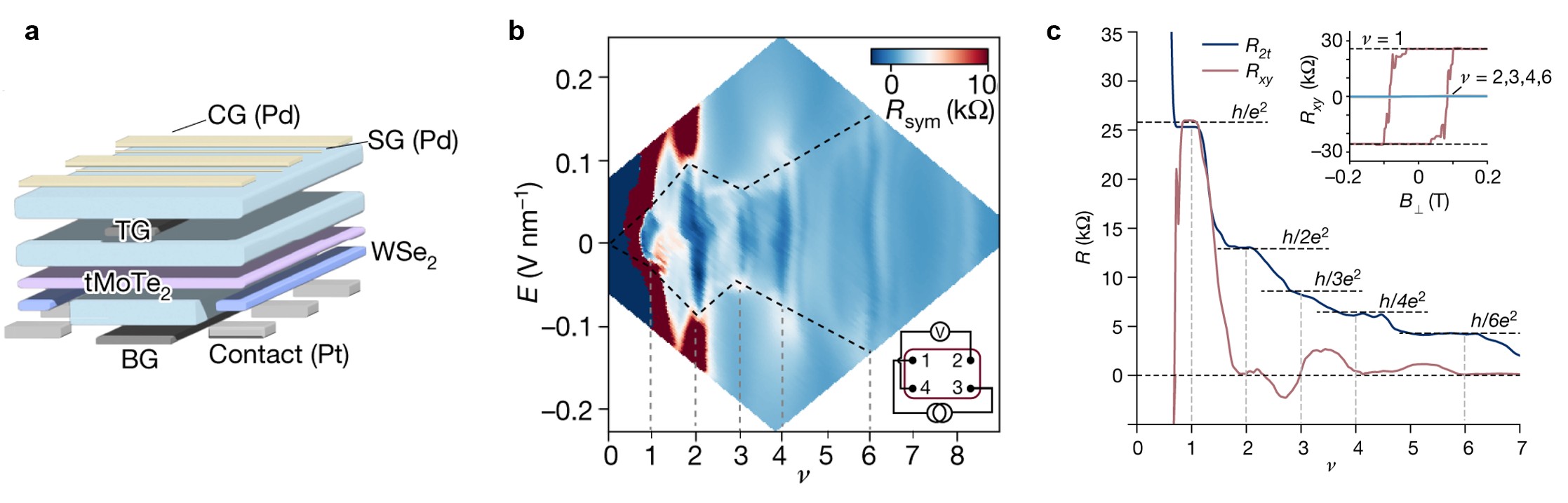}
\caption{\textbf{The fractional quantum spin Hall effect (FQSH) in twisted MoTe$_2$.}
\textbf{a}, Schematic of dual-gated twist MoTe$_2$ devices (2.1°). 
\textbf{b,} Four-terminal resistance $R_{sym}$ versus vertical electric field $E$ and filling factor $\nu$ at 20 mK. $R_{sym}$ is the field symmetric part of the measured resistance at $B = \pm 0.1$ T.
\textbf{c,} Filling factor dependence of two-terminal resistance ($R_{2t}$) and Hall resistance ($R_{xy}$) at $E = 0$ and 20 mK. Inset, $R_{xy}$ shows a magnetic hysteresis and a quantized Hall resistance $h/e^2$ at zero magnetic field for $\nu=1$; the Hall resistance is nearly zero for $\nu=$ 2, 3, 4 and 6. 
Note that Figs.~\textbf{(a-c)} are adapted from Ref.~\cite{kang2024evidence}.
}
\label{FQSH_tMoTe2}
\end{figure}

Among the variety of twist combinations for TMDs, twisted bilayer MoTe$_2$ ($t$-MoTe$_2$) is particularly interesting as it exhibits a rich phase diagram that evolves with the twist angle. Recently, Kang et al.~\cite{kang2024evidence} provided experimental evidence of single and double QSH insulators in $t$-MoTe$_2$, thereby validating theoretical predictions, and further revealed the experimental evidence of a triple QSH phase in this system. In the large-angle regime (approximately 5.08$^\circ$ down to 2.88$^\circ$), $t$-MoTe$_2$ exhibits QSH behavior at $\nu$ = 2 and double QSH behavior at $\nu$ = 4, reflecting a progressive enhancement of multi-helical edge conduction as the moir\'e superlattice flattens. Near 2$^\circ$, the moir\'e potential strongly flattens and isolates the topmost valence bands, enabling robust QSH states driven by this moir\'e potential. In such narrow bands, both double and triple QSH phases can emerge, each edge contributing conductances of $G_0, 2G_0$, and $3G_0$, respectively ($G_0 = e^2/h$, Figs.~\ref{FQSH_tMoTe2}\textbf{b-c}). As the twist angle further decreases, the moir\'e bands become progressively flatter, ultimately yielding multiple flat Chern bands; at 1.89$^\circ$, up to five flat Chern bands with Chern number C= 1 can exist~\cite{xu2025multiple}, underscoring the critical role of moir\'e engineering in tailoring the topological landscape of the system.

Apart from the $t$-MoTe$_2$, Kang et al.~\cite{kang2024double} also identify the similar QSH and double QSH phases in twisted bilayer WSe$_2$ ($t$-WSe$_2$, Figs.~\ref{edge_under_Bfield}\textbf{g-i}). At a twist angle between 2.5$^{\circ}$ and 3.5$^{\circ}$, strong moir\'e potential and weak disorder stabilize a QSH phase at hole filling factor $\nu$ = 2, corresponding to one pair of helical edge states. When the filling factor reaches $\nu$ = 4, the system transitions into a double QSH phase, supporting two pairs of helical edge states (Fig.~\ref{edge_under_Bfield}\textbf{g}). Due to the Ising-type SOC in the two K valleys, the spin-up and spin-down states within each valley are completely split, ensuring the conservation of the spin $S_z$ component. As a result, this phase can also be characterized by nearly quantized edge transport, large nonlocal resistance, and robustness against out-of-plane magnetic fields. 

An alternative platform is the AB-stacked MoTe$_2$/WSe$_2$ hetero-bilayer. A moiré potential is formed due to the lattice mismatch between the two materials. In a $\nu = 1$ experiment, Tao et al.~\cite{tao2024giant} discovered a QAH phase and giant spin-hall effect, indicating the presence of large SOC. Around the same time, Zhao et al.~\cite{zhao2024realization} observed QSH insulator phases in their MoTe$_2$/WSe$_2$ devices when doped to $\nu = 2$. Furthermore, Zhao et al. achieved a topological transition from QSH insulator to QAH insulator in $\nu = 2$ by applying an out-of-plane magnetic field. A QSH insulator where S$_z$ is a good quantum number can be thought of as having two pairs of S$_z$ polarized bands, each of which in isolation is a Chern insulator with opposite Chern numbers ($C = \pm1$). In this scenario, an out-of-plane magnetic field would have the opposite effect on each pair of bands. Under a sufficiently large magnetic field, one pair of bands will invert, resulting in a non-zero Chern number in the overall system. A key experimental challenge in achieving this is that, while a sizable topological gap is required for the observation of robust QSH edge states, a small enough band gap is required for a reasonable magnetic field to invert part of the bands. In Zhao's case, the AB-stacked WSe$_2$ and MoTe$_2$ heterostructure features a spin-valley-layer-locked band structure with a convenient gap size. In the QSH phase, each valley hosts one (spin-polarized) QSH edge. By applying a 2 T magnetic field, the authors were able to induce a band inversion in one valley. In addition, by leveraging the layer-polarized band structure, the authors could tune the size of the QSH gap by applying an out-of-plane displacement field, changing the amount of magnetic field required to achieve such a transition. In fact, a sufficiently large displacement was also observed to cause a QSH insulator to trivial insulator transition.

\section{Fractional QSH and Fractional Chern Insulators}

\begin{figure}
\centering
\includegraphics[width=0.8\textwidth]{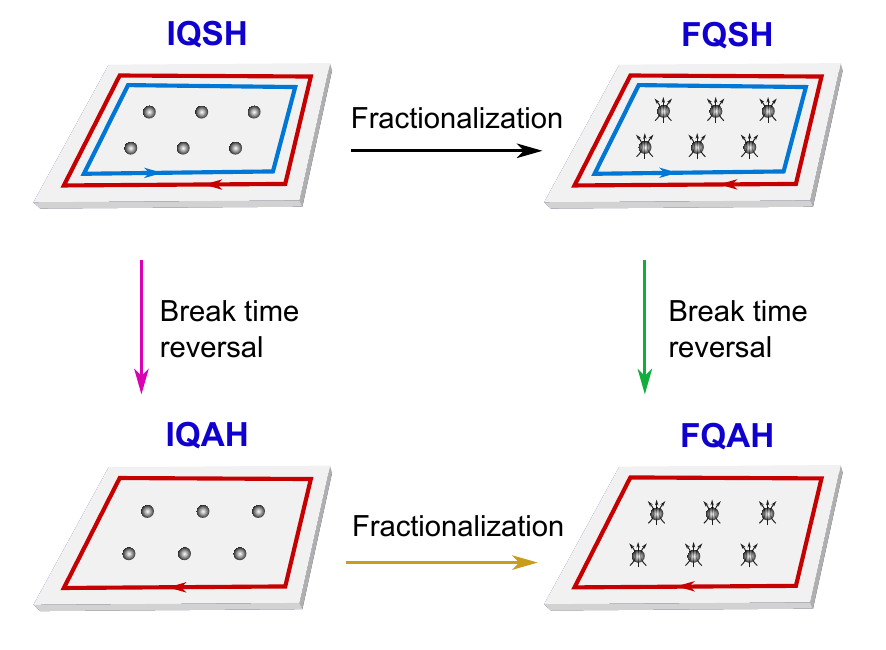}
\caption{\textbf{From QSH to other topological states.} The evolution from QSH to QAH, FQSH and FQAH states by breaking time-reversal symmetry and fractionalization process.}
\label{Transformation}
\end{figure}

QSH systems, under time-reversal symmetry breaking and strong correlations, can give rise to new topological phases (Fig.~\ref{Transformation})~\cite{oh2013complete, liu2008quantum}. This has been most clearly exemplified recently in TMD moiré systems. 
In moiré systems, the presence of flat bands and enhanced density of states gives rise to strong correlation effects, which interact with the underlying topological band structure and lead to a rich landscape of correlated topological phases~\cite{cai2023signatures,zeng2023thermodynamic,xu2023observation,park2023observation,kang2024evidence,park2025ferromagnetism, xu2025interplay, xia2025superconductivity, guo2025superconductivity, xu2025signatures}. Among these, the fractional QSH and fractional Chern insulator has been realized in moiré TMDs~\cite{park2023observation,xu2023observation,kang2024evidence}. The FQSH and FQAH states are characterized by a fractionally quantized Spin Hall conductivity or fractional quantized Hall conductivity in the absence of an external magnetic field, and holds promise for applications in topological quantum computation and the exploration of exotic entangled quantum phenomena.

In general, realizing an FQAH state requires two key ingredients: (i) spontaneous magnetism and (ii) fractionally quantized conduction~\cite{reddy2023fractional,wang2024fractional,xu2024maximally}. Recently, $t$-MoTe$_2$ gained interest due to its distinct electronic structure, strong SOC, and highly tunable electron correlations. While conventional magnetism originates from localized magnetic moments associated with partially filled $d$ or $f$ orbitals, the magnetism in $t$-MoTe$_2$ stems from an imbalance in carrier occupation between the $K$ and $K'$ valleys. SOC locks spin to valley, enabling spontaneous valley and spin polarization at fractional or odd-integer fillings. This valley imbalance results in a net magnetization, detectable via magnetic circular dichroism, and is manifested by hysteresis loops, indicating spontaneous time-reversal symmetry breaking. Furthermore, the moiré-engineered valence bands in $t$-MoTe$_2$ near the Fermi level possess nontrivial Chern numbers and extremely narrow bandwidths, forming flat Chern bands analogous to Landau levels. At partial fillings, these bands are expected to exhibit fractional quantum Hall-like behavior, thereby fulfilling both requirements for realizing FQAH phases: spontaneous magnetism and fractionally quantized conduction.

\begin{figure}
\centering
\includegraphics[width=\textwidth]{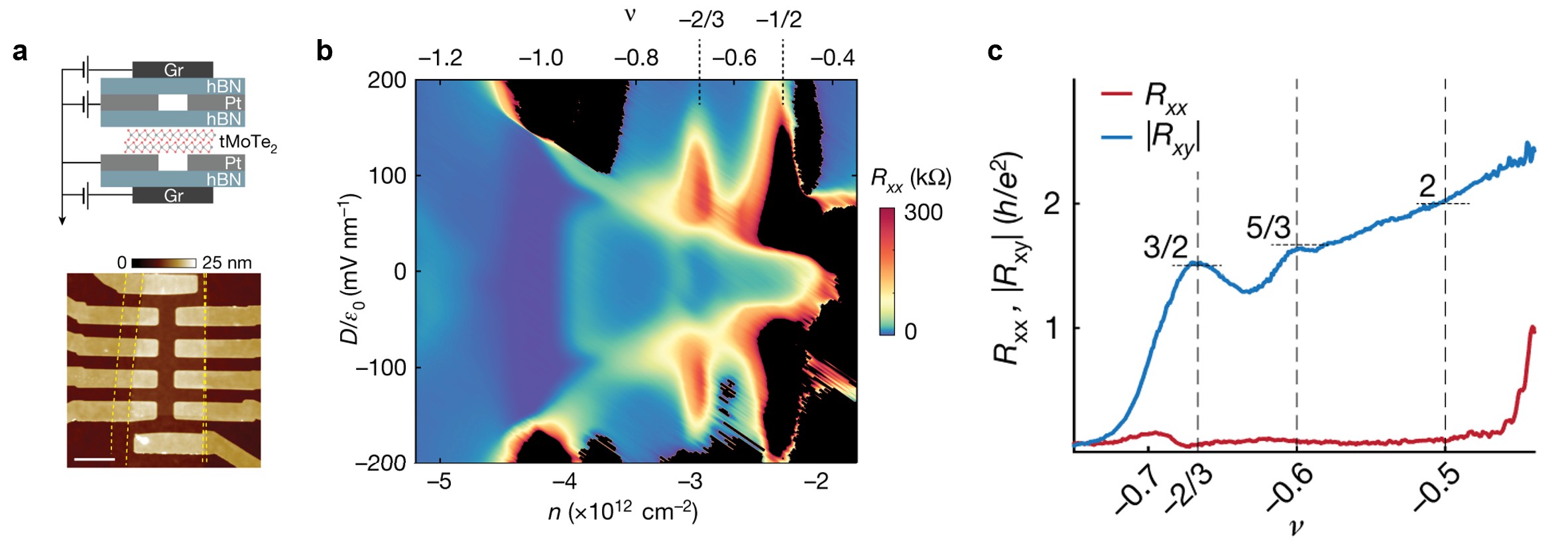}
\caption{\textbf{The fractional quantum anomalous Hall effect (FQAH) in twisted MoTe$_2$.}
\textbf{a}, Top, the device geometry schematic of a twist MoTe$_2$ devices (3.7°). Bottom, atomic force microscopy image during fabrication. Dashed lines indicate the edges of the MoTe$_2$ flakes. Scale bar, 2 $\mu$m.
\textbf{b,} Longitudinal ($R_{xx}$) resistance as a function of electric field ($D/\epsilon_0$) and carrier density ($n$) at 100 mK. 
\textbf{c,} Symmetrized $R_{xx}$ (red) and anti-symmetrized $R_{xy}$ (blue) at $\pm50$ mT versus filling factor $\nu$ at $T = 100$ mK.
Note that Figs.~\textbf{(a-c)} are adapted from Ref.~\cite{park2023observation}.
}
\label{FQAH_tMoTe2}
\end{figure}

Cai et al.~\cite{cai2023signatures} first reported signatures of FQAH states at a hole filling of $\nu = -2/3$ in $t$-MoTe$_2$ at relatively large twist angles around $\theta \sim 3^{\circ}$ (Fig.~\ref{FQAH_tMoTe2}), using optical probes. Magneto-circular dichroism was used to detect spontaneous magnetism, while trion luminescence measurements were employed to identify gapped incompressible states. By mapping these states as a function of carrier density and magnetic field, a Landau fan diagram was constructed, showing that the incompressible states follow the Středa formula—indicative of their topological nature. The topological character of the $\nu = -2/3$ state was further supported by thermodynamic measurements of electronic compressibility and Landau fan diagram~\cite{zeng2023thermodynamic}, and unambiguously confirmed by transport experiments, which revealed fractionally quantized Hall conductance at both $\nu = -2/3$ and $\nu = -3/5$~\cite{park2023observation,xu2023observation}.
Shortly thereafter, Lu et al.~\cite{lu2024fractional,lu2025extended} also observed FQAH states in five-layer and four-layer rhombohedral graphene/BN superlattices.

The fractional QSH represents another type of topological quantum state arising from strong electron correlations. It is characterized primarily by two distinct features: (i) the system exhibits fractionally quantized spin Hall conductance; and (ii) the excitations carry fractionalized charges. Recent transport experiments have reported evidence for a fractional QSH state in $t$-MoTe$_2$ at a filling factor of $\nu = -3$~\cite{kang2024evidence} (Fig.~\ref{FQSH_tMoTe2}). In this regime, each edge was found to contribute a fractional longitudinal conductance of $G = \frac{3}{2}G_0$ per edge, while the Hall conductance remains zero, indicating the presence of fractionalized helical edge channels. However, the precise nature of this state remains under investigation, as magneto-circular dichroism measurements indicate apparent time-reversal symmetry breaking~\cite{kang2025time}.

\section{From QSH to Nonlinear Anomalous Hall}

\begin{figure}
\centering
\includegraphics[width=\textwidth]{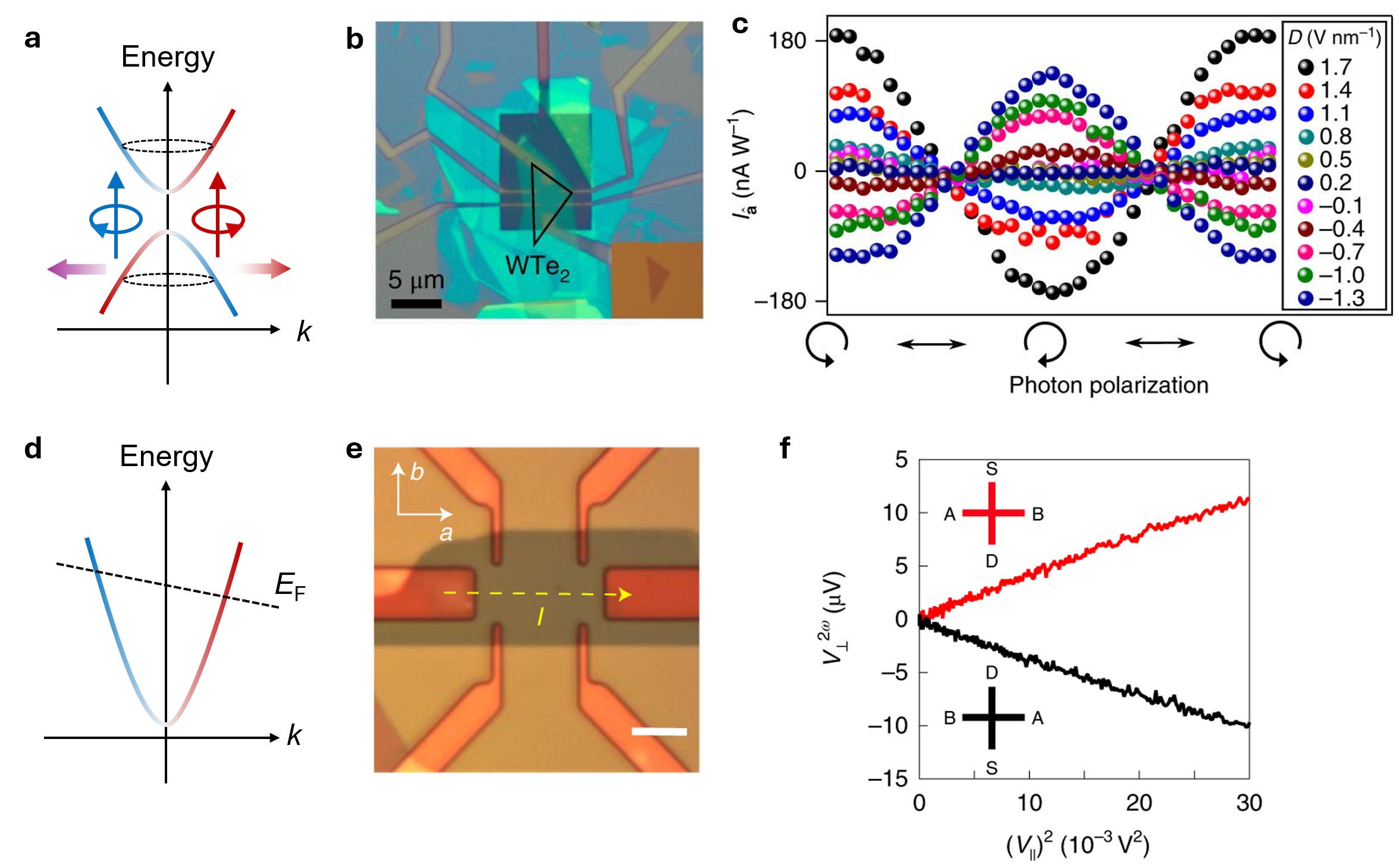}
\caption{\textbf{Experimental observation of Berry curvature dipole effects.}
\textbf{a,} The interband Berry curvature dipole effect.
\textbf{b,} An image of a monolayer WTe$_2$ device with dual gates for measuring the circular photo-galvanic effect. 
\textbf{c,} The measured photocurrent plotted against photon polarization. The different colors represent different applied displacement fields.
\textbf{d,} The intraband Berry curvature dipole effect.
\textbf{e,} An image of a few-layer WTe$_2$ device with hall-bar geometry electrodes for measuring the nonlinear Hall effect. 
\textbf{f,} Measured nonlinear Hall voltage ($V_{\perp}^{2\omega}$) plotted against the square of the applied voltage ($V_{\parallel}$). 
Note that Figs.~\textbf{(b-c)} are adapted from Ref.~\cite{xu2018electrically}, Figs.~\textbf{(e-f)} from Ref.~\cite{kang2019nonlinear}.
}
\label{BCD}
\end{figure}

Band inversion in QSH systems is a source of large Berry curvatures, but their existence and distribution are constrained by symmetry. Under time-reversal symmetry, the Berry curvature is odd in momentum and integrates to zero over the Brillouin zone. If inversion symmetry is additionally present, the Berry curvature vanishes identically at every $k$-point. However, certain QSH systems exhibit sufficiently low crystal symmetry to support a finite Berry curvature dipole (BCD)—a dipolar momentum-space distribution of Berry curvature. This dipole can drive nonlinear optical and transport phenomena, including the circular photogalvanic effect (CPGE)~\cite{de2017quantized} and the NLH~\cite{sodemann2015quantum} (Fig.~\ref{BCD}).

The BCD-induced CPGE was first demonstrated by Xu et al.~\cite{xu2018electrically} in monolayer WTe$_2$. The monolayer is approximately inversion-symmetric, suppressing the BCD. However, a vertical displacement field strongly breaks inversion symmetry and generates a sizable dipole. Due to a mirror symmetry, the BCD is constrained to lie perpendicular to the mirror plane, and reversing the displacement field flips the dipole direction. Circularly polarized light selectively excites carriers in momentum space according to the Berry curvature distribution, producing a photocurrent whose direction reverses with the BCD—a key experimental observation~\cite{xu2018electrically}. 

Bilayer WTe$_2$, though not a QSH insulator, exhibits strong inversion symmetry breaking and a pronounced BCD due to band hybridization. Ma et al.~\cite{ma2019observation} observed a time-reversal-symmetric nonlinear Hall effect in this system, originating from Fermi-surface BCD. By applying an in-plane AC current along the BCD direction, they detected a transverse voltage oscillating at twice the driving frequency—a signature of the second-order Hall response. In contrast to the topologically trivial bilayer WTe$_2$, bilayer TaIrTe$_4$ has been theoretically predicted to be a QSH insulator~\cite{guo2020quantum}, although this remains to be experimentally verified. The system also exhibits strong inversion symmetry breaking, making it an ideal platform for investigating BCD effects. Kumar et al.~\cite{kumar2021room} reported room-temperature microwave rectification in 20~nm-thick TaIrTe$_4$ devices, demonstrating the presence of a BCD-enabled nonlinear Hall effect in this material. It will be interesting to explore whether the bilayer exhibits enhanced BCD-related phenomena, and whether it exhibits unique signatures associated with the topological regime, if experimentally realized. 

Few-layer WTe$_2$ systems, such as bilayer, trilayer and four layer WTe$_2$ in the T$_d$ stacked structure, all exhibit ferroelectricity. Interestingly, Wang and Qian~\cite{Wang2019nhe} discovered a subtle difference between even-layer and odd-layer WTe$_2$, where the ferroelectric configurations of even-layer WTe$_2$ are related via an out-of-plane mirror operation followed by an in-plane shift and the configurations of odd-layer WTe$_2$ are related via inversion operation. Consequently, the BCD and shift dipole will reserve their direction and the NLH voltage will switch sign in odd-layer WTe$_2$ only~\cite{Wang2019nhe}, while such reversal is absent in the even-layer structures. This predicted ferroelectric NLH effect was experimentally observed by Xiao et al.~\cite{Xiao2020berrymemory} in trilayer and four layer WTe$_2$, establishing the concept of layer-parity-selective Berry curvature memory and opening an avenue towards exploring hidden stacking orders and their effect on nonlinear quantum transport and photocurrent~\cite{Wang2019nonlinearphotocurrent}.

\section{Interfacing QSH with other vdW layers}

The integration of topological insulators with 2D magnetic materials offers a highly promising platform for realizing novel quantum phenomena, including the QAH effect~\cite{chang2013experimental,deng2020quantum,liu2020robust} and nonreciprocal magnetoelectric responses~\cite{yasuda2016large,lv2018unidirectional,tokura2018nonreciprocal,fan2019unidirectional,li2024observation}, as well as the formation of skyrmions and the emergence of the Dzyaloshinskii-Moriya interaction (DMI)~\cite{wu2020neel}. These systems enable the exploration of unique quantum states and interactions that are crucial for advancing spintronics and quantum computing. Additionally, the proximity effect between topological insulators and superconducting materials has proven to be a powerful tool for investigating exotic quantum states, such as topological superconductivity and Majorana bound states~\cite{alicea2012new,li2019exploring}, which hold potential for scalable fault-tolerant quantum computing. This synergy between topological insulators, magnetic materials, and superconductors opens new avenues for discovering and harnessing quantum effects in next-generation technologies.

The impact of magnetic proximity on topological edge states has been explored in two recent studies, where the helical edge states of monolayer WTe$_2$ were coupled with the 2D magnetic materials, CrI$_3$ and Cr$_2$Ge$_2$Te$_6$ ~\cite{zhao2020magnetic,li2022proximity,kao2025unconventional}. Zhao et al.~\cite{zhao2020magnetic} investigated the coupling between the monolayer WTe$_2$ and the layered antiferromagnetic insulator CrI$_3$ in a heterostructure. Due to spin-momentum locking, the helical edge states in WTe$_2$ are highly sensitive to magnetic perturbations. Their study revealed that the edge conductance in WTe$_2$ is strongly modulated by the magnetization state of adjacent CrI$_3$ layers, with suppression occurring due to an exchange field-induced gap. Furthermore, the nonlinear edge conductance exhibits a pronounced nonreciprocal response, with a current asymmetry of up to 100$\%$ between the two antiferromagnetic ground states of CrI$_3$. This effect is attributed to spin-flip scattering assisted by magnons, which selectively suppress current flow based on the magnetization direction.

Li et al.~\cite{li2022proximity} studied the coupling between the monolayer WTe$_2$ and a ferromagnetic insulator Cr$_2$Ge$_2$Te$_6$ heterostructure. They demonstrated proximity-induced ferromagnetism in WTe$_2$ through the observation of the anomalous Nernst effect, anomalous Hall effect, and anisotropic magnetoresistance. The edge states in the magnetized WTe$_2$ exhibited partially spin-polarized transport, distinct from ideal 1D helical or chiral edge channels. These findings observe a strong magnetic proximity effect in the heterostructures with 2D magnetic materials, demonstrating a viable approach for tuning topological edge states through magnetic order. They also highlight the potential of vdW heterostructures in advancing spintronics and quantum computing by exploiting the interplay between topology and magnetism for low-power, nonreciprocal devices. 

The proximity effect at the 2D topological insulators-magnetic materials interface also induces a strong DMI, essential for stabilizing Néel-type skyrmions. Wu et al.~\cite{wu2020neel} fabricated WTe$_2$/Fe$_3$GeTe$_2$ heterostructures and observed chiral spin textures through transport measurements, revealing the topological Hall effect below 100 K, an indication of skyrmion formation. Lorentz transmission electron microscopy directly imaged a Néel-type skyrmion lattice and stripe-like magnetic domains. The estimated DMI energy at the interface, $\sim$1.0 mJ$\cdot$m$^{-2}$, is comparable to values in heavy metal/ferromagnet systems. This heterostructure advances 2D magnetism and skyrmion physics, offering potential for high-density, low-power spintronic memory and logic devices.

Kao et al.~\cite{kao2022deterministic} demonstrate deterministic switching of perpendicular magnetization using unconventional spin-orbit torques (SOTs) in the low-symmetry quantum material WTe$_2$, in proximity to the vdW ferromagnet Fe$_3$GeTe$_2$ (Fig.~\ref{apps}\textbf{c}). Similar works~\cite{liu2023field,bainsla2024large,zhang2023room} are also demonstrated in the TaIrTe$_4$ and conventional ferromagnets, such as Py (Ni$_{80}$Fe$_{20}$), cobalt iron boron (CoFeB), etc. 
The researchers experimentally achieved field-free magnetization switching by leveraging the out-of-plane antidamping torque in WTe$_2$, which arises due to its unique crystal symmetry. Unlike conventional spin-source materials, WTe$_2$ enables out-of-plane spin polarization when a charge current is applied along its low-symmetry axis, facilitating the switching of Fe$_3$GeTe$_2$'s perpendicular magnetization without an external magnetic field (Fig.~\ref{apps}\textbf{d}). The proximity effect between WTe$_2$ and Fe$_3$GeTe$_2$ enhances spin-to-charge conversion, with the out-of-plane antidamping torque playing a crucial role in the deterministic switching process. Numerical simulations further confirm that the observed switching is driven by unconventional SOTs in WTe$_2$, highlighting the potential of low-symmetry quantum materials for next-generation spintronic applications.

L\"upke \textit{et al.}~\cite{lupke2020proximity} reported the coexistence of superconductivity and QSH edge states in a van der Waals heterostructure composed of monolayer WTe$_2$ and superconducting NbSe$_2$. The superconductivity from NbSe$_2$ results in a proximity-induced superconducting gap in WTe$_2$ while preserving its QSH edge states, thus creating a potential 1D topological superconductor. The ability to induce superconductivity in WTe$_2$ without electrostatic doping, combined with the high critical temperature of NbSe$_2$, offers an experimental advantage for studying the interplay between superconductivity and topological edge states.

\section{Potential applications of QSH materials}

As a prototypical topological electronic phase, the QSH effect exhibits unique electronic and spintronic functionalities that distinguish it from conventional semiconducting electronics. As highlighted earlier, QSH materials host topologically protected 1D edge channels, which can function as novel interconnects with robust conduction properties. Additionally, the spin-momentum locking inherent to these edge states offers opportunities for spintronic applications, enabling efficient spin current generation and manipulation. Below, we explore several emerging avenues for leveraging the distinctive electronic, spin, and wave-function properties of QSH materials in practical applications. In particular, we examine the potential of utilizing the nonlinear Hall response for high-frequency electronics and microwave energy harvesting. Furthermore, we discuss how the QSH systems could open pathways for realizing topologically protected quantum computing schemes.

\subsection{Utilizing the spin-momentum locked edge conduction}

In a conventional semiconductor field-effect transistor, a metal-to-insulator transition achieves the ``on'' and ``off'' states. By applying a displacement field through the control gate, the relative Fermi level in the channel can be tuned between inside the conduction or valence band and inside the band gap. A similar device could be made with a QSH insulator. By applying a displacement field, we can achieve on-off switching of the topological edge states with a topological phase transition. Due to the robustness against scattering in the helical edge states, such a device would, in theory, have the advantage of near-dissipationless conduction in the ``on'' state.

\begin{figure}
\centering
\includegraphics[width=\textwidth]{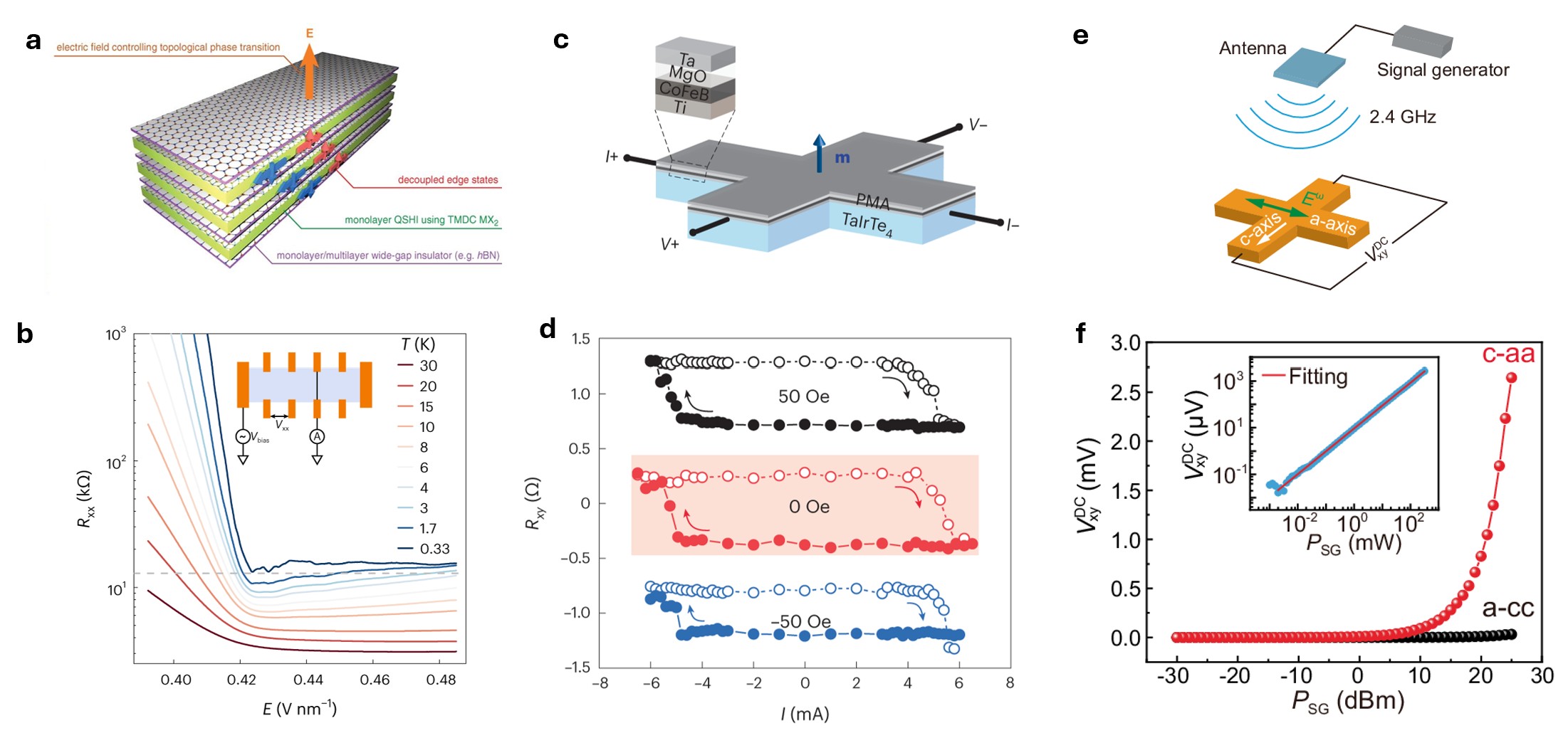}
\caption{\textbf{Applications of QSH materials}
\textbf{a,} A schematic of the design of a topological transistor. Here a single pair of field-effect gates switches multiple channels with the applied displacement fields. \textbf{b,} Measured resistance in an AB-stacked MoTe$_2$/WSe$_2$ device plotted against applied displacement field. Different color lines represent different temperatures. Larger applied displacement field places the device in a QSH insulator phase, as indicated by the near-quantized resistance plateau at low temperature, whereas smaller applied displacement field places the device in a trivial insulator phase. \textbf{c,} A schematic of a device for current-driven perpendicular magnetization-switching measurements. \textbf{d,} The Hall resistance versus the current from -6 to 6 mA. The different colors indicate the different applied external magnetic field, with the red curve measured with no applied magnetic field.  \textbf{e,} A schematic device for rectifying a wireless signal using the nonlinear Hall effect. \textbf{f,} Measured nonlinear Hall effect plotted against the power of the applied wireless signal. Different colors indicate different measurement configurations. The inset is a log-log plot of the c-aa direction.
Note that Fig.~\textbf{(a)} are adapted from Ref.~\cite{Qian2014Quantum}, Fig.~\textbf{(b)} from Ref.~\cite{zhao2024realization}, Figs.~\textbf{(c-d)} from Ref.~\cite{liu2023field}, Figs.~\textbf{(e-f)} from Ref.~\cite{cheng2024giant}.
}
\label{apps}
\end{figure}

Qian et al.~\cite{Qian2014Quantum} proposed the possibility of such a device in 1T$^{\prime}$-MX$_2$ (Fig.~\ref{apps}\textbf{a}). Due to the already low symmetry of 1T$^{\prime}$-MX$_2$, applying an out-of-plane displacement field further breaks inversion symmetry, increases Rashba type SOC, and causes the originally spin-degenerate bands to split. When enough displacement field is applied, a band inversion for each spin-polarized band occurs at the avoided crossings of the original band inversion, causing a transition to a trivial insulator phase. However, due to the large displacement field required (on the order of $2$ V/nm) to trigger the topological transition, the experimental realization of such a transition has been elusive. The exact strength of Rashba-type splitting in response to the applied displacement field depends on the microscopic details of electron distribution and can therefore vary from material to material.

Zhao et al.~\cite{zhao2024realization} observed a displacement field-tuned QSH insulator to trivial insulator transition in an AB-stacked MoTe$_2$/WSe$_2$ moir\'e device (Fig.~\ref{apps}\textbf{b}). In their system, the before-band-inversion conduction band and valence band are layer polarized. Applying an out-of-plane electric field has the straightforward effect of modifying the potential energy of the layers, thereby directly tuning the energy of the conduction band and valence band relative to each other. Due to the relatively small band gap to layer separation ratio, only about $0.425$ V/nm of electric field was required for the trivial insulator to QSH insulator transition to occur.

Other than using them for conduction, the chiral, spin-momentum locked edge states in a QSH insulator also have potential applications in spintronics. A crucial technology required to build spintronics is the interconversion between spin and charge. In theory, a DC edge current should be entirely spin-polarized. Indeed, Kondou et. al. have shown that topological insulators can produce very high charge to spin conversion ratios~\cite{kondou2016fermi}. Furthermore, Araki \textit{et al.}~\cite{araki2020dynamical} theoretically predicted that QSH insulator edge states can exhibit efficient spin-to-charge conversion when a magnetic field induces a gap in the edge spectrum. This prediction remains to be confirmed experimentally. On the experimental side, Bainsla \textit{et al.}~\cite{bainsla2024large} demonstrated that the strong SOC in bulk TaIrTe$_4$ gives rise to a pronounced spin-orbit torque effect, enabling magnetization switching of an adjacent ferromagnet without the need for an external magnetic field~\cite{liu2023field} (Fig.~\ref{apps}\textbf{c, d}).

\subsection{Energy harvesting with QSH materials}

The NLH effect~\cite{sodemann2015quantum} arises naturally in QSH insulators with sufficiently low crystal symmetry. The band inversion responsible for QSH topology often results in a large Berry curvature dipole, which drives the NLH effect. Onishi \textit{et al.}~\cite{onishi2024high} have predicted that rectifiers based on the NLH effect could outperform conventional semiconductor-based technologies in converting electromagnetic radiation into usable energy. This is particularly promising for low-power and high-frequency regimes (e.g., THz), where traditional technologies exhibit sharply reduced efficiency~\cite{lu2014wireless, hemour2014towards, okba2019compact, zhang2021terahertz, shen2024nonlinear, cheng2024giant}.

Among ambient electromagnetic energy harvesting methods, two primary technologies stand out: solar cells and RF rectennas. Solar cells serve as the benchmark and operate efficiently in the visible spectrum. However, their external quantum efficiency decreases substantially in the near-infrared regime, particularly beyond 1100~nm, due to the silicon band gap. Rectennas, antenna and rectifier combinations, extend the usable spectrum to longer wavelengths, including the RF domain (Figs.~\ref{apps}\textbf{e-f}). Ambient RF energy is sufficient to power certain classes of sensors~\cite{zhang2020opportunities}. Despite advances in wireless power transfer, including improved antenna design, impedance matching, and coupling efficiency, the rectifier remains the limiting component, with state-of-the-art efficiencies below 1\% in the high-frequency regime~\cite{citroni2022progress}. This inefficiency stems from conventional rectifiers’ reliance on energy barriers or band gaps, which suppress nonlinear response at low input powers~\cite{hemour2014towards}.

In contrast, NLH-based rectifiers offer a fundamentally different mechanism. Because rectification arises from a gapless intraband process, they exhibit nonlinear response even at arbitrarily small input power~\cite{onishi2024high}, due to their inherently parabolic current-voltage characteristics. Moreover, the transverse nature of the NLH current ensures that the rectified signal is orthogonal to the driving electric field, resulting in zero Joule heating ($\vec{J} \cdot \vec{E} = 0$). Limited only by the electron scattering time, NLH rectification is expected to operate across a broad frequency range from DC to THz.

Despite these theoretical advantages, the practical implementation of NLH-based energy harvesters requires further experimental validation and quantitative analysis to assess efficiency, scalability, and device integration. Initial demonstrations of NLH rectification in the GHz range~\cite{kumar2021room, zhang2019two, cheng2024giant, lu2024nonlinear, kumar2024quantum, makushko2024tunable, zhou2025room} offer promising proof-of-concept. Given their predicted performance~\cite{onishi2024high,zhang2021terahertz}, frequency coverage, and structural simplicity, NLH rectifiers are strong candidates for the next generation of ambient energy harvesting technologies.

\subsection{Topological quantum computing}
There has been tremendous interest in finding platforms that host Majorana zero modes (MZM). An MZM is a special quasi-particle excitation in 2D condensed matter systems. They are named after the famous Majorana fermion conjectured as a fundamental particle~\cite{majorana2020symmetric, moore1991nonabelions, kitaev2001unpaired, yazdani2023hunting}. These quasiparticles are non-abelian in the sense that a collection of them has a topologically degenerate ground state; upon exchanging two of these quasiparticles by ``braiding'' them around each other, one goes from one ground state to another  (e.g. from $\ket0$ to $(\ket0+\ket1)/\sqrt2$), instead of merely gaining a phase factor in the case of conventional fermions and bosons or even abelian anyons. MZMs could potentially allow us to store each qubit in one or more spatially separated pairs of MZMs, such that no local environmental disturbance can couple to the qubit and scramble the information (decohere). This could lead to quantum computers with significantly fewer physical qubits required to achieve a logical qubit compared to existing approaches~\cite{fu2010electron, vijay2016teleportation, karzig2017scalable, schrade2018parity}. While controlled braiding may be difficult to achieve, the same effect can be achieved with measurement operations~\cite{bonderson2008measurement, vijay2016teleportation}. 

The existence of these non-abelian quasiparticles was first proposed in 1991 in the special $5/2$ fractional quantum Hall state, where quasiparticle excitations are believed to exist in the form of pairs of composite fermions~\cite{moore1991nonabelions}. Kitaev proposed later in 2001 that they could also exist in p-wave superconductors, which can host Majorana fermions localized at the vortex core and chiral Majorana mode propagating along sample edges. Recently, signatures of $p$-wave superconductivity in rhombohedral trilayer graphene~\cite{han2024signatures} and in moir\'e TMDs~\cite{xu2025signatures,xu2025chiral} have been reported.

In a superconductor, a Cooper pair can be broken to form excited states known as Bogoliubov quasiparticles at finite energy. These quasiparticles are a superposition of electron and hole states~\cite{bardeen1957microscopic}. A MZM is a special type of bogoliuvbov quasiparticle with an even mixture of an electron and a hole (without charge), and with zero energy. By bringing a conventional pairing potential to conventional topological gapless edge modes~\cite{fu2008superconducting, zhang2008px+, qi2010chiral, oreg2010helical, zhang2013majorana, nilsson2008splitting}, the Hamiltonian of the exotic p-wave superconductor can be mimicked. For example, MZMs can be created at junctions of superconductor proximitized QSH edge states and magnetically gapped QSH edge states~\cite{nilsson2008splitting, fu2009josephson}. Topological phases such as QAH or QSH provides robust edge states for engineering MZMs without the need of a large magnetic field.

A quantum computer built with MZMs alone is unfortunately not universal, and requires ancillary qubits prepared in special states with no topological protection to achieve universal computation. Theorists have proposed more exotic non-abelian quasi-particles known as parafermions that could potentially provide full-fledged universal computation with topological protection~\cite{freedman2002modular}. A proposed realization of these parafermions makes use of a fractional QSH edge proximitized with a superconductor~\cite{cheng2012superconducting}.

\section{Challenges and opportunities}

\subsection{Achieving robust helical edges}

Despite the tremendous progress made over the past few years, significant challenges remain in the field of QSH systems.

While both WTe$_2$ and TaIrTe$_4$ exhibit compelling evidence of a transition between ballistic and diffusive transport regimes---with the ballistic regime showing the expected quantized conductance of $2e^2/h$---the quantization length in these materials remains relatively short, typically less than $1\,\mu\text{m}$. This limitation is intrinsically linked to the nature of the QSH effect, which is protected by time-reversal symmetry. As a result, any scattering event that breaks time-reversal symmetry can suppress edge conduction. This stands in stark contrast to the quantum Hall effect, where edge states are protected by gauge symmetry and are significantly more robust against disorder.

There are a few things that can be done. For instance, creating atomically sharp edges could significantly improve the edge state coherence. In this context, quasi-1D systems are particularly promising, as clean edges can be naturally produced during exfoliation along the 1D chain direction. Such systems could provide a platform for studying QSH physics in highly controlled environments, potentially leading to new insights and applications.

On the other hand, recent experiments on TMD moiré suggest the possibility of achieving longer quantization lengths, on the order of micrometers~\cite{zhao2024realization,kang2024evidence,kang2024double}. This improvement may arise from the long moiré length scale in these systems, as well as the potential conservation of $S_z$ (spin projection along the $z$-axis), which could protect edge quantization independently of time-reversal symmetry. Exploring this mechanism further could provide a promising route for enhancing the robustness of QSH edge states.

\subsection{Flatband QSH materials and fractional $ \mathcal{Z}_2 $ insulators}

Despite the evidence of fractional QSH states in twisted MoTe$_2$~\cite{kang2024evidence}, many questions remain unresolved, both theoretically and experimentally. Unlike most other QSH systems~\cite{Qian2014Quantum,liu2017van,fei2017edge,wu2018observation,Tang2024DualQSH}, where time-reversal symmetry is the key protective mechanism, these systems are protected by $ S_z $ conservation. The nature of these systems is particularly intriguing due to the conservation of $ S_z $, which allows for both time-reversal symmetric and asymmetric versions of the QSH phase.  The QSH phase in such systems can be described using spin Chern numbers and can be viewed as two copies of the QAH effect. Similarly, the fractional QSH phase can be interpreted as two copies of the fractional QAH effect.

A compelling direction for future research is the exploration of flatband QSH systems that may host fractional topological states, particularly in the absence of $S_z$ conservation. In such systems, the topological phase cannot be characterized by spin Chern numbers or decomposed into independent quantum Hall-like components. Instead, these phases are classified by two gapless edge states of fractional anyons related by time reversal symmetry, rendering them fundamentally distinct from conventional quantum Hall states. For instance, twisting intrinsically $\mathcal{Z}_2$ materials such as WTe$_2$ or TaIrTe$_4$ may enhance electronic correlations within the QSH bands and facilitate the emergence of such exotic fractionalized phenomena.

\subsection{Novel Control of QSH}

Strain provides an alternative tuning knob for driving topological phase transitions. It is particularly effective in systems where the conduction and valence bands possess opposite bonding characters (bonding vs.~antibonding) and form a direct bandgap. Applying uniaxial strain along the bonding axis shifts the relative energies of these bands, potentially inducing a band inversion and realizing a topological transition~\cite{liu2014manipulating}. Such strain-driven transitions have been experimentally demonstrated in three-dimensional crystals~\cite{mutch2019evidence,lin2021visualization,liu2024controllable}. With the emergence of advanced strain-tuning platforms for 2D materials~\cite{liu2024continuously,hwangbo2024strain}, similar tuning strategies may soon be implemented in QSH insulators.

Another emerging control knob is the twist angle, particularly relevant for moiré systems. Several techniques have recently been developed to enable \textit{in situ} control of interlayer twist angles at cryogenic temperatures. One such method is the quantum twisting microscope (QTM)~\cite{inbar2023quantum,Birkbeck2025quantum,klein2024imaging}, in which one layer is mounted on an atomic force microscope
(AFM) tip and the other on a substrate. By rotating the tip, the interlayer twist angle can be dynamically tuned. Concurrently, QTM operates as a scanning tunneling probe that enables momentum-resolved tunneling into moiré quantum states. Leveraging coherent multi-path interference and momentum conservation, QTM visualizes the energy dispersion of electronic states (e.g., Dirac cones, flat bands) and reveals quantum phenomena inaccessible to conventional STM techniques. Applying QTM to twisted moiré QSH systems would be highly valuable for probing the twist-angle dependence of topological phases and directly visualizing their electronic structures. Another platform, the microelectromechanical system (MEMS)-based on-chip multi-degree-of-freedom control technique developed by Tang et al.~\cite{tang2024chip}, provides similar capabilities for dynamic manipulation of interlayer registry with multiple independent tuning parameters. These advanced platforms open new avenues for controlling quantum phases and exploring QSH phenomena in moiré systems with continuous angle control.

\noindent\textbf{Acknowledgments:} 

Q.M., J.T., and T.S.D. acknowledge support from the Air Force Office of Scientific Research (grants FA9550-22-1-0270 and FA9550-24-1-0117), the Office of Naval Research (grant N00014-24-1-2102), and the Alfred P. Sloan Foundation. V.B. acknowledges support from NSF ITE-2345084. C.W. and X.Q. gracefully acknowledge the support from the National Science Foundation under grants DMR-1753054 and DMR-2103842. N.M. and Y.Z. are supported by the Max Planck Partner lab on quantum materials, and the National Science Foundation Materials Research Science and Engineering Center program through the UT Knoxville Center for Advanced Materials and Manufacturing (DMR-2309083).

\noindent\textbf{Author contributions:} All authors wrote the manuscript together.

\noindent\textbf{Competing interests:} The authors declare that they have no competing interests.

\newpage

\clearpage
\newpage


%

\end{document}